\documentclass[twocolumn]{aastex631}

\newcommand{\hph}{$76.7^{+15.5}_{-6.7}$~km~s$^{-1}$~Mpc$^{-1}$}
\newcommand{\hphsp}{$75.7^{+8.1}_{-5.5}$~km~s$^{-1}$~Mpc$^{-1}$}
\newcommand{\htd}{$71.8^{+9.2}_{-8.1}$~km~s$^{-1}$~Mpc$^{-1}$}

\newcommand{\hphn}{$76.7^{+15.5}_{-6.7}$}
\newcommand{\hphspn}{$75.7^{+8.1}_{-5.5}$}
\newcommand{\htdn}{$71.8^{+9.2}_{-8.1}$}
\newcommand{\kms}{km~s$^{-1}$}
\newcommand{\kmsM}{\kms~Mpc$^{-1}$}
\usepackage{amsmath}
\newcommand{\Msol}{\hbox{M$_\odot$}}

\begin{document}

\title{SN H0pe: The First Measurement of $H_0$ from a Multiply-Imaged Type Ia Supernova, Discovered by JWST}

\author[0000-0002-2282-8795]{Massimo Pascale}
\affiliation{Department of Astronomy, University of California, 501 Campbell Hall \#3411, Berkeley, CA 94720, USA}

\author[0000-0003-1625-8009]{Brenda L.~Frye}
\affiliation{Department of Astronomy/Steward Observatory, University of Arizona, 933 N. Cherry Avenue, Tucson, AZ 85721, USA}

\author[0000-0002-2361-7201]{Justin D.R. Pierel}
\affiliation{Space Telescope Science Institute, 3700 San Martin Drive, Baltimore, MD 21218, USA}

\author[0000-0003-1060-0723]{Wenlei Chen}
\affiliation{Department of Physics, Oklahoma State University, 145 Physical Sciences Bldg, Stillwater, OK 74078, USA}

\author[0000-0003-3142-997X]{Patrick L. Kelly}
\affiliation{Minnesota Institute for Astrophysics, University of Minnesota, Minneapolis, MN 55455, USA}

\author[0000-0003-3329-1337]{Seth H. Cohen} 
\affiliation{School of Earth and Space Exploration, Arizona State University,
Tempe, AZ 85287-1404, USA}

\author[0000-0001-8156-6281]{Rogier A. Windhorst} 
\affiliation{School of Earth and Space Exploration, Arizona State University, Tempe, AZ 85287-1404, USA}

\author[0000-0002-6124-1196]{Adam G. Riess}
\affiliation{Space Telescope Science Institute, 3700 San Martin Drive, Baltimore, MD 21218, USA}
\affiliation{Department of Physics and Astronomy, Johns Hopkins University, Baltimore, MD 21218 USA}

\author[0000-0001-9394-6732]{Patrick S. Kamieneski}
\affiliation{School of Earth and Space Exploration, Arizona State University,
Tempe, AZ 85287-1404, USA}

\author[0000-0001-9065-3926]{Jos\'e M. Diego}
\affiliation{Instituto de Fisica de Cantabria (CSIC-C). Avda. Los Castros s/n. 39005 Santander, Spain}

\author[0000-0002-7876-4321]{Ashish K. Meena}
\affiliation{Department of Physics, Ben-Gurion University of the Negev, P. O. Box 653, Be’er-Sheva, 84105, Israel}

\author[0000-0001-7148-6915]{Sangjun Cha}
\affiliation{Department of Astronomy, Yonsei University, 50 Yonsei-ro, Seoul 03722, Korea}

\author[0000-0003-3484-399X]{Masamune Oguri}
\affiliation{Center for Frontier Science, Chiba University, Chiba 263-8522, Japan}
\affiliation{Department of Physics, Graduate School of Science, Chiba University, 1-33 Yayoi-Cho, Inage-Ku, Chiba 263-8522, Japan}

\author[0000-0002-0350-4488]{Adi Zitrin}
\affiliation{Department of Physics, Ben-Gurion University of the Negev, P. O. Box 653, Be’er-Sheva, 84105, Israel}

\author[0000-0002-5751-3697]{M. James Jee}
\affiliation{Department of Astronomy, Yonsei University, 50 Yonsei-ro, Seoul 03722, Korea}
\affiliation{Department of Physics and Astronomy, University of California, Davis, One Shields Avenue, Davis, CA 95616, USA}

\author[0000-0002-7460-8460]{Nicholas Foo}
\affiliation{School of Earth and Space Exploration, Arizona State University, Tempe, AZ 85287-1404, USA}

\author[0009-0001-7446-2350]{Reagen Leimbach}
\affiliation{Department of Astronomy/Steward Observatory, University of Arizona, 933 N. Cherry Avenue, Tucson, AZ 85721, USA}

\author[0000-0002-6610-2048]{Anton M. Koekemoer} 
\affiliation{Space Telescope Science Institute,
3700 San Martin Drive, Baltimore, MD 21218, USA}

\author[0000-0003-1949-7638]{C.~J.~Conselice} 
\affiliation{Jodrell Bank Centre for Astrophysics, Alan Turing Building,
University of Manchester, Oxford Road, Manchester M13 9PL, UK}

\author[0000-0003-2091-8946]{Liang Dai}
\affiliation{Department of Physics, University of California, 366 Physics North MC 7300, Berkeley, CA. 94720, USA}

\author[0000-0002-4163-4996]{Ariel Goobar}
\affiliation{The Oskar Klein Centre for Cosmoparticle Physics, Department of Physics, Stockholm University, SE-10691 Stockholm, Sweden} 

\author[0000-0003-2445-3891]{Matthew R. Siebert}
\affiliation{Space Telescope Science Institute,
3700 San Martin Drive, Baltimore, MD 21218, USA}

\author[0000-0002-7756-4440]{Lou Strolger}
\affiliation{Space Telescope Science Institute,
3700 San Martin Drive, Baltimore, MD 21218, USA}

\author[0000-0002-9895-5758]{S. P. Willner}
\affiliation{Center for Astrophysics \textbar\ Harvard \& Smithsonian, 60 Garden Street, Cambridge, MA 02138, USA}

\begin{abstract}
The first James Webb Space Telescope ({\it JWST}) Near InfraRed Camera (NIRCam) imaging in the field of the galaxy cluster PLCK G165.7+67.0 ($z=0.35$) uncovered a Type Ia supernova (SN~Ia) at $z=1.78$, called ``SN H0pe." Three different images of this one SN were detected as a result of strong gravitational lensing, each one traversing a different path in spacetime, thereby inducing a relative delay in the arrival of each image. Follow-up {\it JWST} observations of all three SN images enabled photometric and rare spectroscopic measurements of the two relative time delays. Following strict blinding protocols which oversaw a live unblinding and regulated post-unblinding changes, these two measured time delays were compared to the predictions of seven independently constructed cluster lens models to measure a value for the Hubble constant, $H_0=$~\htdn~\kmsM. The range of admissible $H_0$ values predicted across the lens models limits further precision, reflecting the well-known degeneracies between lens model constraints and time delays. It has long been theorized that a way forward is to leverage a standard candle, but this has not been realized until now. For the first time, the lens models are evaluated by their agreement with the SN absolute magnifications, breaking  degeneracies and producing our best estimate, $H_0=$~\hphsp\,. This is the first precision measurement of $H_0$ from a multiply-imaged SN~Ia and only the second from any multiply-imaged SN.
\end{abstract}


\section{Introduction} \label{sec:intro}

Determination of the current expansion rate of the Universe (the Hubble Constant, $H_0$) across a wide range of redshifts provides a fundamental test of the standard cosmological model, $\Lambda$CDM. While this model succeeds in reproducing an abundance of observed cosmological phenomena, disagreement between independent measurements of $H_0$ at early and late times in the Universe raises doubts about the model's reliability. 
This tension primarily results from two independent precision methods for measuring $H_0$: Type Ia supernovae (SNe~Ia) using the local-distance-ladder approach and cosmic mircowave background (CMB) observation in conjunction with the $\Lambda$ cold dark matter ($\Lambda$CDM) model.
Confirming or resolving the $H_0$ discrepancy is crucial for fundamental physics including the effects of neutrinos, the nature of dark energy, and the spatial curvature of the Universe.
 
For the local distance-ladder, the `Supernova $H_0$ for the Equation of State' (SH0ES) team's  \citep{Riess2022} 
calibration of distances to nearby Cepheid variable stars in multiple ways, followed by calibration of SNe~Ia which share host galaxies with Cepheids,
yields $H_0 = 73.29 \pm 0.090$~km~s$^{-1}$~Mpc$^{-1}$ \citep{Murakami2023}. In contrast, CMB measurements \citep{Planck2020b} from the \textit{Planck} satellite favor $H_0 = 67.4 \pm 0.5$~km~s$^{-1}$~Mpc$^{-1}$ assuming a flat $\Lambda$CDM cosmology. This apparent ${>}5\sigma$ disagreement between two premier methods is the basis of what is known as the `Hubble tension'. 
Examining other independent $H_0$ measurements can delineate the underlying discrepancy. Local-distance-ladder measurements using Hubble Space Telescope (HST) imaging of the `tip of the red-giant branch' (TRGB) stars to calibrate SNe~Ia 
from the Carnegie-Chicago Hubble Program \citep[CCHP,][]{Freedman2019}, the Extragalactic Distance Database \citep[EDD,][]{Anand22}, and  Comparative Analysis of TRGBs \citep[CATs,][]{Scolnic23} yield a range in $H_0$ of 70--73~km~s$^{-1}$~Mpc$^{-1}$. Through comparison of measured colors to measured surface-brightness fluctuations (SBF) calibrated on TRGB and Cepheids, \cite{Blakeslee21} found $H_0= 73.3 \pm 0.7 \pm 2.4$~km~s$^{-1}$~Mpc$^{-1}$, while \cite{Garnavich2023} found $74.6 \pm 0.9 \pm 2.7$~km~s$^{-1}$~Mpc$^{-1}$ by adding SNe~Ia as the top rung. In a `single-rung' distance ladder, the Maser Cosmology Project \citep[MCP,][]{Pesce20} combined six maser distances and peculiar velocities to find $H_0=73.9 \pm 3.0$~km~s$^{-1}$~Mpc$^{-1}$.  Another single-rung method uses time variability between the multiple images of strongly gravitationally lensed quasars. The Time Delay COSMOgraphy (TDCOSMO) collaboration leveraged a joint measurement across seven lensed quasars to find $H_0=74.2^{+1.6}_{-1.6}$~km~s$^{-1}$~Mpc$^{-1}$ assuming power-law mass profiles for the lens and a flat $\Lambda$CDM cosmology, and $H_0=73.3^{+5.8}_{-5.8}$~km~s$^{-1}$~Mpc$^{-1}$ using free-form mass profiles (\citealt{Shajib2023}; see also \citealt{Millon2020}, \citealt{Wong2020}, and \citealt{Birrer2020}). This growing catalog of independent measurements has already provided insights into the Hubble tension \citep{Verde2023}, and new measurements at intermediate redshifts $1<z<10$ are imperative in elucidating when in cosmic history this discrepancy occurs.

\cite{Refsdal1964} first recognized that strong gravitational lensing of SNe enables another independent method for measuring $H_0$. The `strong' regime of gravitational lensing refers to the deflection of the light of a distant background object (star, galaxy, or QSO) by a foreground galaxy or cluster of galaxies such that the image of a single object is observed at multiple positions on the sky. 
For each observed image of an ``image system," the background object's 
light traverses a different path length, resulting in different travel times for photons to each image. Transient events within the lensed object (e.g., a supernova or quasar variability) enable measurement of the time delay between multiple images. The time delay, together with a model of the lensing mass distribution, yields the timed delay distance to the source object, a quantity inversely proportional to $H_0$. 

Although initially envisioned for SNe, inference of $H_0$ through strong-lensing time delays was first applied to multiply-imaged quasars lensed by single foreground galaxies \citep{Suyu2010}. The technique, known as time-delay cosmography, was not realized for SNe until the discovery of `SN Refsdal', a multiply-imaged Type II SN found in the Massive Cluster Survey (MACS) J1149.6+2223 cluster field \citep{Kelly2015}. Crucially, strongly-lensed SNe in galaxy-cluster lenses provide an independent route for measuring $H_0$ whose systematics are distinct from the distance-ladder, quasar-time-delay cosmography, and CMB approaches \citep{Suyu2023}. SN Refsdal was observed as four multiple images in an Einstein cross around a single elliptical-galaxy lens (a `galaxy--galaxy' lens). A fifth image with an $\sim$1~year relative time delay was predicted \citep{Sharon2015,Diego2016,Oguri2015,Treu2016,Grillo2016} and later observed \citep{Kelly2016}. The relatively long time delay enabled SN Refsdal to be the first lensed SN for which meaningful constraints of $H_0$ could be inferred, and the nature of the galaxy-cluster-scale lens gave these constraints distinct systematics from previous galaxy--galaxy lens methods \citep{Grillo2018,VegaFerrero2018}. The constraints and their systematics were later quantified by \cite{Grillo2020}. The most robust constraints to date from SN Refsdal were provided by \cite{Kelly2023a}, who inferred $H_0 = 64.8^{+4.4}_{-4.3}$~km~s$^{-1}$~Mpc$^{-1}$ or $H_0 = 66.6^{+4.1}_{-3.3}$~km~s$^{-1}$~Mpc$^{-1}$ depending on the set of lens models used. \cite{Liu2024} and \cite{Grillo2024} further explored the systematics, employing lens modeling approaches designed explicitly for time-delay cosmography.

On the heels of the discovery of SN Refsdal, several other lensed supernova have been observed. These include other Type II SN, such as the $z\sim3$ SN in the Abell 370 cluster field which allowed detailed studied of shock cooling in a red supergiant SN \citep{Chen2022}, as well as the SN~Ia `iPTF16geu' at $z=0.409$ \citep{Goobar2017} was the first SN~Ia to have multiple images observed. It was followed by `SN Requiem'  at  $z=1.95$ \citep{Rodney2021}, and `SN Zwicky' at $z=0.3554$ \citep{Goobar2023,Pierel2023}. Both iPTF16geu and Zwicky were galaxy-scale lenses, whose $\sim$day-long time delays limited inference of $H_0$ to $\gtrsim$40\% precision \citep{Dhawan2020,Pierel2023}. By contrast, the galaxy-cluster-scale lens of SN Requiem has a nearly decade-long time delay, which will yield a precision measurement of $H_0$ in 2037 when the SN counterimage is predicted to appear \citep{Rodney2021}. In the same host galaxy as SN Requiem, a second SN, SN `Encore', was discovered, marking the first occurrence of a lensed galaxy producing multiple observed SNe \citep{Pierel2024}.  Across all of these SNe, SN Refsdal is the only one to provide precision $H_0$ constraints to date. 

SN H0pe is a triply-imaged SN~Ia discovered in the galaxy-cluster field PLCK G165.7+67.0 (`G165', $z=0.348$). It is a massive lensing galaxy cluster \citep[$M_{\rm tot} = (2.6\pm0.3) \times 10^{14}$~\Msol,][]{Frye2019, Pascale2022a}, that induces a relatively long $\sim$100 day time delay between the first and third image, offering only the second-ever opportunity for precision SN time-delay cosmography and the first  with a SN~Ia. The three supernova multiple images were discovered in  {\it JWST}/NIRCam imaging in the lensed, NIR-bright host galaxy `Arc~2', which itself is triply imaged \citep{Frye2023an}. \cite{Polletta2023} first measured the Arc~2 redshift as $z=1.783 \pm 0.002$. The SN H0pe discovery paper \citep{Frye2023b} used followup  {\it JWST}/NIRSpec spectroscopy to identify SN H0pe as a SN~Ia, to confirm the SN host galaxy redshift to higher precision ($z=1.7834\pm0.0005$), and to spectroscopically-confirm a coincident galaxy overdensity at the redshift of SN H0pe. 
The known spectroscopic evolution of SNe~Ia allowed \cite{Chen2023H0pe} to determine the SN phases in the three multiple images of SN~H0pe and thereby 
measure relative time delays of $-122.3^{+43.7}_{-43.8}$~days and $-49.3^{+12.2}_{-14.7}$~days between images a--b and c--b respectively.
Image b is the last to arrive and consequently shows the earliest phase of the SN, which was near peak brightness at the time of discovery.
This is the first time spectroscopy has provided useful time delay constraints. 
\cite{Chen2023H0pe} also classified SN H0pe as a normal SN~Ia. 
\cite{Pierel2023H0pe} used three epochs of  {\it JWST} NIRCam six-band imaging to photometrically measure relative image time delays of $-116^{+10.8}_{-9.3}$~days and $-48.3^{+3.6}_{-4.0}$~days for images a--b and c--b respectively.

Independently through the photometric and spectroscopic datasets respectively, \cite{Pierel2023H0pe} and \cite{Chen2023H0pe} use the standard-candle nature of the SN to constrain the absolute magnifications induced by strong lensing. Absolute magnification measurements have long been thought to break degeneracies among mass models inherent in $H_0$ measurements from strong-lens time delays \citep{Falco1985,Kolatt1998, Oguri2003,Schneider2013}, yet have previously been unavailable for time-delay cosmography. For example in the case of the cluster-lensed SN Refsdal, the inferred $H_0$ was found to vary across twenty-three different lens mass models despite all satisfying the same lensing constraints. However, since $H_0$ was found to strongly correlate with the predicted magnifications, magnification measurements could have broken this degeneracy \citep{Liu2024}.
SN H0pe provides the first opportunity to harness the standard candle for time-delay cosmography.

This paper presents the first precision measurement of $H_0$ from time-delay cosmography of a strongly-lensed SN~Ia. The outline is as follows. Section~\ref{sec:theory} summarizes the underlying theory of time-delay cosmography. Section~\ref{sec:obs} describes the {\it HST} and {\it JWST} observations that led to SN H0pe's discovery and subsequent analysis. Sections~\ref{sec:phot} and~\ref{sec:spec} briefly summarize the photometric light-curve fitting and spectroscopic age-dating analyses. Section~\ref{sec:models} details the construction of the seven contributing lens models. Section~\ref{sec:bayes} outlines the process through which $H_0$ was derived from measured observables. Section~\ref{sec:blind} outlines the guidelines for the blinded inference. Section~\ref{sec:results} presents the results of the $H_0$ inference. Section~\ref{sec:disc} discusses the unique advantages provided by the SN~Ia, the robustness of the various inference approaches, and the overall error budget. Finally, section~\ref{sec:concl} discusses future work and how SN H0pe fits into the broader picture of time-delay cosmography.

Throughout this analysis, we enforced a blinding between the seven independent lens models, the photometrically-derived time delays, and the spectroscopically-derived time delays from each other. Changes made after the unblinding and the impacts of each change are explicitly discussed in Section \ref{sec:blind}.

\section{Time-Delay Cosmography} \label{sec:theory}
Here we briefly outline how the time delay, $\Delta t_{i,j}$, between any two lensed images $i$ and $j$ of a multiply-imaged source depends on the positions of the observed multiple images with respect to the source, the local lensing potential, and the assumed cosmological model. For an individual source at angular position $\beta$ with a corresponding lensed image observed at angular position $\theta$, the time delay can be expressed as:
\begin{align} \label{eq:td1}
    t (\boldsymbol{\theta},\boldsymbol{\beta}) = \frac{1+z_\mathrm{l}}{c}\frac{D_\mathrm{l} D\mathrm{s}}{D_\mathrm{ls}}[\frac{1}{2}(\boldsymbol{\theta} - \boldsymbol{\beta})^2 - \psi(\boldsymbol{\theta})]\quad,
\end{align}
where $z_\mathrm{l}$ is the redshift of the lensing cluster, $\psi(\boldsymbol{\theta})$ is the lensing potential at the observed image position $\boldsymbol{\theta}$, and $D_\mathrm{l}$, $D_\mathrm{s}$, and $D_\mathrm{ls}$ are the angular diameter distances to the lens, the source, and between the lens and source respectively. If instead we measure the time delay between any two images of a set of multiple images:
\begin{align} \label{eq:tdij}
    \Delta t_{i,j}(\boldsymbol{\theta_i},\boldsymbol{\theta_j}) = D_{\Delta t}\Delta \tau_{i,j}(\boldsymbol{\theta_i}, \boldsymbol{\theta_j}, \boldsymbol{\beta})\quad,
\end{align}
where $\tau$ is the Fermat potential, $\tau(\boldsymbol{\theta},\boldsymbol{\beta}) = \frac{1}{2}(\boldsymbol{\theta} - \boldsymbol{\beta})^2 - \psi(\boldsymbol{\theta})$ \citep{Schneider1985,Blandford1986}. $D_{\Delta t}$ is the so-called `time delay distance', $D_{\Delta t} = \frac{1+z_l}{c}\frac{D_\mathrm{l} D_\mathrm{s}}{D_\mathrm{ls}}$, and is dependent on the angular diameter distances and hence the cosmological model \citep{Refsdal1964,Schneider1992,Suyu2010}. $D_{\Delta t}$ is inversely proportional to $H_0$, while being only weakly dependent on other cosmological parameters \citep{Bonvin2017}. Hence, if the time delay between any two multiple images is known, as well as the redshifts of the lens and source, then a model of the lensing potential is necessary to constrain $H_0$.

In practice a lens model can be constructed from lensing observables, such as the observed positions of multiply-imaged galaxies or information of the foreground lensing galaxies, and assumes a fiducial value of $H_0$, $H_0^{\rm fid} = 70$~\kmsM, to predict a fiducial time delay $\Delta t_{i,j}^{\rm fid}$. Following equation~\ref{eq:td1}, the true measured time delay for any image pair can be related to this lens model fiducial time delay by rescaling $H_0$ while holding all other cosmological parameters fixed, in turn inferring $H_0$:
\begin{align} \label{eq:rescale}
    \Delta t_{i,j}^{\rm pred}(H_0) = \Delta t_{i,j}^{\rm fid} \times \frac{70~ {\rm km}~{\rm s}^{-1}~{\rm Mpc}^{-1}}{H_0}\quad.
\end{align}

An alternative approach can be accomplished by using measured time delays as input lens model constraints and freely optimizing for all cosmological parameters \citep[e.g.,][]{Grillo2020,Grillo2024}, which has the advantage of probing degeneracies between $H_0$ and other cosmological parameters under different cosmologies. However functionality for this was not available to all lens models in this work, and the use of the time delays as input constraints for the lens models would have conflicted with the blinding goals of this work, as multiple lens model teams had constructed G165 lens models prior to the discovery of SN H0pe. Additionally, this would have added time delay ratio constraints to these models, which we instead reserved as a blinded prediction to be incorporated into the final weighting scheme (see Section \ref{sec:bayes}). Finally, we point out that when imposing a standard flat $\Lambda$CDM cosmology, the lens model time delay is expected to vary $\lesssim 1\%$ with respect to $\Omega_m$ across a typical range $\Omega_m \in [0.2,0.4]$ for the redshift of SN H0pe. While it is possible to explore parameter degeneracies under other cosmologies through such an approach (e.g., open $\Lambda$CDM), this is outside the intended scope of this work.

\section{Observations} \label{sec:obs}
\begin{figure*}[t!]
\includegraphics[scale=0.25]{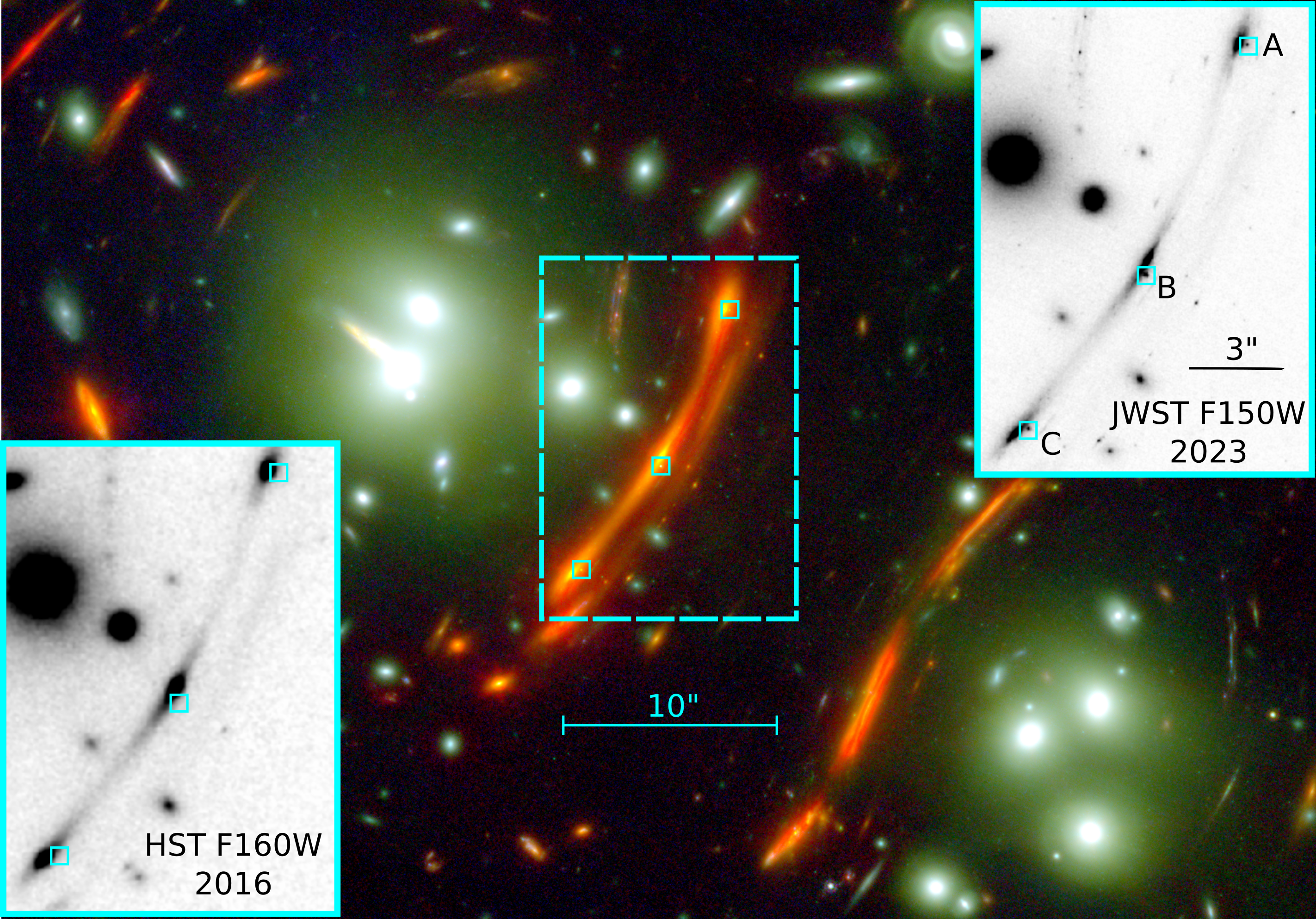}
\caption{{\it JWST}/NIRCam color image in the central region of G165. Insets show closeup of the boxed region depicting the three multipole images of the host galaxy Arc 2 prior to SN H0pe in \textit{HST} WFC3/IR F160W imaging from 2016 (lower left) and during its appearances in \textit{JWST}/NIRCam F150W imaging from 2023 (upper right). The three images of the Type Ia SN H0pe appear only in the 2023 images, and their positions are marked by the blue boxes, where the `A', `B', and `C' refer to images $a$, $b$, and $c$. North is up and east is to the left.
}
\label{fig:discovery}
\end{figure*}

The discovery paper for SN H0pe \citep{Frye2023b} supplied details regarding the {\it JWST} imaging and spectroscopic data sets.  In brief, SN H0pe was initially discovered in NIRCam images from the PEARLS GTO program \citep[PI: Windhorst, PID: 1176;][]{Windhorst2023}. The discovery images were taken on 31 March 2023 in eight filters. The SN was recognized by comparing the F150W PEARLS NIRCam imaging to WFC3/IR F160W HST images taken on 30 May 2016 (Figure~\ref{fig:discovery}). The discovery prompted a  {\it JWST} disruptive Director's Discretionary Time (DDT) program which acquired two additional epochs of NIRCam imaging in six filters (PI: Frye, PID: 4446) on 22 April 2023 and 9 May 2023. Photometric analysis of the galaxy cluster field was presented by \cite{Frye2023b}, and photometric light-curve fitting of the SN was presented by \cite{Pierel2023H0pe}. 

\begin{figure*}[t!]
\includegraphics[scale=0.51]{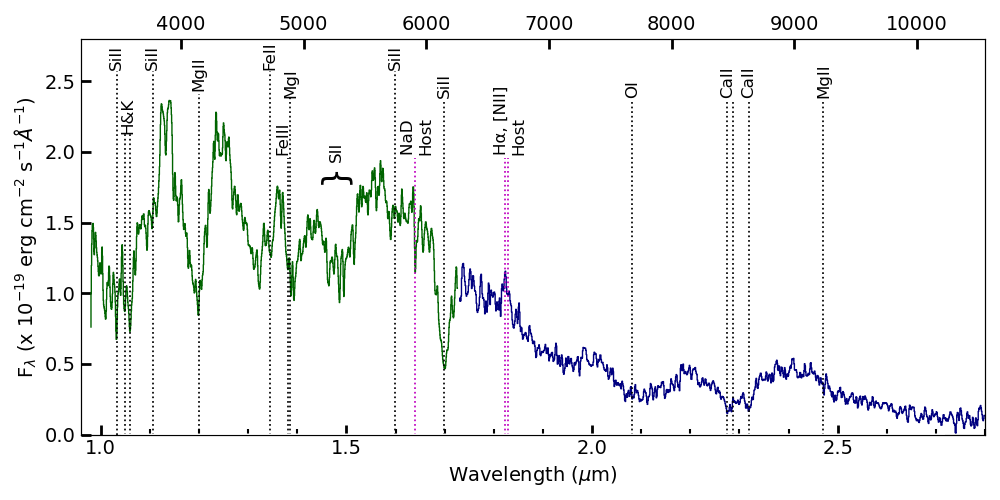}
\caption{High signal-to-noise NIRSpec spectrum of SN H0pe. The G140M (green) and G235M (blue) grating spectra are shown.  Some prominent line features are identified, such as the [\ion{Si}{2}] line blueshifted to rest-frame $\sim6150$~\AA\ that is a requiste line feature of a Type Ia supernova. A few features are also traced to the SN host galaxy, such as NaD, H$\alpha$, [\ion{N}{2}]$\lambda\lambda$6548,6584 (magenta dotted-line). The observed wavelength is given on the lower abscissa and the rest wavelength in the frame of the host galaxy is provided on the upper abscissa.}
\label{fig:sn_spec}
\end{figure*}

The DDT program also acquired {\it JWST}/NIRSpec Micro-shutter assembly (MSA) spectra \citep{Frye2023b,Chen2023H0pe} in the PRISM, G140M, and G235M gratings targeting the SN host galaxy and the two brightest SN multiple images. 
The spectra, taken on 22 April, confirmed the Ia type from the identification of the requisite blueshifted 
[\ion{Si}{2}]$\lambda$6355 feature, Ca~H\&K, and other absorption-line features commonly found in Type Ia SNe
(Figure~\ref{fig:sn_spec}).

\section{Photometric Time Delays and 
Magnifications} \label{sec:phot}
The light-curve evolution of SNe~Ia is well-studied \citep[e.g.,][]{Hsiao2007,Pierel2021}, 
such that precision measurements of the time delay can be made by fitting measured SN colors across multi-epoch imaging. Given that there are three epochs of NIRCam imaging, and each epoch contains three SN multiple images, the SN evolution is photometrically sampled at nine points in time. Here we briefly summarize the light-curve fitting analysis of \cite{Pierel2023H0pe}. The halo of the bright galaxy host, a significant source of noise, is reconstructed and subtracted through simultaneous fitting of its surface brightness profile and cluster lens model with the \texttt{GLEE} software package \citep[][details in Caminha et al., in prep.]{Halkola2006,Suyu2010}. 
The photometry of the SN multiple images is then fit simultaneously to BayeSN SN~Ia models \citep{Mandel2022,Grayling2024}. Time delays are inferred by fitting a common set of light curve parameters to each of the SN multiple images. The magnification ratios are estimated by normalizing the light curve to image $b$, and then computing the ratios of images $a$ and $c$ relative to $b$. Finally, the absolute magnifications are estimated by comparing the peak F277W flux ($\sim$rest-frame Y band) predicted by the BayeSN models, to the population of field SNe\,Ia.

Statistical uncertainties of the inferred time delays and magnifications are estimated by bootstrapping the photometric uncertainties assuming Gaussian errors. Systematic uncertainties resulting from the BayeSN model itself are produced in a similar manner. Systematic uncertainties from chromatic microlensing are estimated by examining the range of fits to a set of 1000 simulated SN H0pe light curves with variable realizations of the foreground microlensing. Systematics from millilensing by dark matter subhalos in the foreground cluster lens are achromatic, and applied directly as a correction to the distribution of inferred magnifications. Finally, the $\sim 0.1$ mag intrinsic scatter of SN~Ia absolute magnitudes \citep{Scolnic2018} is applied to the magnification distribution.

\section{Spectroscopic time delays and Magnifications} \label{sec:spec}

Spectra from multiple images of a strongly-lensed SN obtained during a single observing epoch enable the measurement of time delays via spectroscopic-based phase analysis. Previously, the lack of quality simultaneously-obtained spectra have precluded such an analysis. 
Since SN H0pe was discovered from the PEARLS NIRCam Epoch 1 observation when its multiple images were near peak brightnesses, the rapid acquisition of spectra in a ``disruptive" {\it JWST}
DDT program using the {\it JWST} NIRSpec's MSA enabled the identification of the SN as Type Ia through the detection of the requisite blueshifted [\ion{Si}{2}]$\lambda$6155 absorption-line feature.

The identification of SN H0pe as a SN~Ia allowed for more straightforward template fitting to determine the phases between multiple images using spectroscopic methods. Below, we describe the process of measuring time delays through spectroscopic age-dating from NIRSpec data. We refer to \citet{Chen2023H0pe} for details.

Extraction of the SN spectrum from the NIRSpec data is performed using custom software designed to cope with background subtraction, host galaxy contamination, and bad pixels (e.g., cosmic rays). The error contribution from these potential systematics is included in the overall uncertainties, along with other sources of error such as slit loss and flux calibration. The relative delays between the multiple images' emitting times are estimated by determining their SN phases (days post-B-band peak brightness) through a joint Markov Chain Monte Carlo (MCMC) fit of the SN spectra against various spectroscopic templates \citep[e.g.][]{Hsiao2007}. The magnification is estimated by comparing the rest-frame $Y$-band magnitude of the best fit spectral models with that of an unmagnified SN~Ia at $z=1.783$. Systematics from microlensing, millilensing, and the $\sim 0.1$ mag intrinsic scatter in SN Ia absolute magnitudes are addressed precisely as described in the photometric light-curve fitting section outlined previously.

\section{Lens Modeling} \label{sec:models}
To obtain the cluster lens model-predicted time delays and magnifications, seven different teams agreed to construct independent lens models for this analysis. All lens modelers were supplied with the same input constraints. These consist of 21 multiple image systems (comprising 109 multiple images, 15 of which pertain to the SN and features within its host galaxy). 
Five multiple image systems have spectroscopically-confirmed redshifts, which corresponded to 8 multiple images with spectroscopic data (1a, 1b, 1c, 2a, 2c, 5a, 8c, and 9c).
The image-system and cluster-member catalogs are public \citep{Frye2023b}. 

The multiple image systems were selected predominantly by eye based on similar colors and morphology and informed by spectroscopic, photometric,  and lens model-predicted redshift constraints. All lens modeling teams participated in the development and vetting of the image system catalog, which was unanimously agreed upon.

Cluster members were determined via the 1.6~$\mu$m bump in color--color space from NIRCam photometry \citep{Sawicki2002}. Candidates were additionally vetted through spectroscopic redshifts (where available) and a standard red-sequence cut \citep{Gladders2000}. This cut removed two small galaxies situated near the SN host arc (R.A., Decl.\ = 11:27:15.84, +42:28:32.41 and 11:27:15.62, +42:28:30.462). Photometric redshifts also indicate that these are in the background at $0.5<z<1$. Additionally a bright, early-type galaxy (R.A., Decl.\ = 11:27:06.70, +42:27:50.46)  in the NIRCam parallel field $>$2\arcmin\ from the apparent position of the SN host is in the foreground at $z\sim0.28$. These galaxies were included in the catalog to the lens modelers to allow them to be optimized separately from the cluster members \citep[e.g.,][]{Raney2020,Acebron2022}. NIRCam imaging with $m_{\rm lim} = 28$--29 (depending on filter)  reveals no other galaxies that may contribute significant line-of-sight structure to the lens, but low-surface-brightness perturbers below the detection limit of the NIRCam images cannot be ruled out. The final catalog consisted of 161 cluster members, of which 21 were spectroscopically confirmed (following the criterion of \citealt{Pascale2022a}).

We emphasize that each lens model was created without knowledge of the photometric or spectroscopic time delay and magnification measurements, and the results of any given model were also blinded from the other modeling teams. All lens models assume a fiducial flat cosmology $\Omega_m=0.3$, $\Omega_\lambda = 0.7$, and $H_0^{\rm fid} = 70$~\kmsM. The time delays and magnification predictions from the lens models are measured at the lens model-predicted positions of the multiple images rather than the observed positions.

We briefly summarize the general methods of lens models included in this analysis, the assumptions made by each, and the expected systematics. A full discussion of each lens model can be found in Appendix~\ref{app:lens_models}. The seven lens models in this analysis are broadly split into two categories. Parametric models (models 1, 2, 3) assume the mass distribution is entirely characterized as the superposition of analytic mass density profiles. These models typically impose large-scale profiles to describe the cluster-scale dark matter halo, and smaller halos to describe each galaxy, which include priors based on the galaxy brightness or morphology \citep{Zitrin2015,Oguri2010,Jullo2007}. Non-parametric models (models 4 and 6), on the other hand, make fewer assumptions on the profile of the cluster-scale dark matter, and allow a flexible grid to describe the mass distribution \citep{Diego2005,Sendra2014}. Some models (models 5 and 7) are so-called `semi-parametric', and may make use of both analytic profiles and more flexible non-parametric distributions. Parametric models benefit from a small number of free parameters and mass distributions which more closely mirror the galaxy configuration of the cluster, while non-parametric methods are given greater freedom in the shape of the underlying mass distribution and include a number of regularization parameters and stopping criteria to avoid overfitting and unphysical solutions \citep{Diego2005, PonenteDiego2011, Cha2022}. 
 
One of the contributing models, model 4, leverages weak gravitational lensing to construct their lens model \citep{Cha2023}. This applies shape distortions of background galaxies in the galaxy cluster field as constraints on the cluster mass distribution, and extends far outside the 
locus of strong-lensing observables in the cluster center \citep[e.g.,][]{Kaiser1993, Hoekstra2000,Jee2006,Bradac2005}. By incorporating weak-lensing into strong-lensing models, it is possible to further constrain the steepness of the mass profile which directly impacts the predicted time delay \citep{Bradac2005,Jee2006,Zitrin2015}. 

Lens models may also differ in how they optimize to estimate their posteriors, that is whether they minimize the separation of lensed multiple images traced back to the source plane or the separation of predicted and observed lensed multiple images in the image plane. The two approaches should in principle perform similarly \citep{Keeton2010}, but source-plane optimization is generally much faster computationally. While there are concerns that source-plane optimization may be less accurate than image-plane optimization in practice \citep[e.g.,][]{Keeton2001}, strong lens modeling in the Frontier-Fields era has shown that both produce quality lens models \citep[e.g.,][]{Meneghetti2017}. In the case of SN Refsdal, \cite{Kelly2023a} successfully applied a sample consisting of both image- and source-plane optimization models infer $H_0$. This work similarly includes lens models which follow both approaches. We find the plane of optimization may in one modeling approach have a non-negligible impact on the resulting time delay and magnification posteriors, which is detailed in Appendix \ref{app:spvsip}. For the case of this model, the image-plane version was chosen over the source-plane, as it was inherently more robust (see Section \ref{sec:blind}.

Lens model diversity broadly accounts for systematics regarding how the lensing mass and hence potential is parametrized. However, a number of systematics are expected to affect each lens model similarly. These include the thin-sheet mass assumption \citep{Falco1985,Birrer2016}, line-of-sight structure \citep{Dalal2005,Acebron2022,Raney2020}, multiplane lensing \citep{McCully2014}, millilensing from dark matter subhalos \citep{Dai2020, Gilman2020}, and the effects of large-scale structure not captured within the field of view \citep{Wong2011}.

Beyond providing a uniform set of lensing inputs, a number of decisions were left up to the expertise of each individual team (e.g, the plane of optimization, uncertainty of the lensed image positions, priors on fitted parameters describing analytic mass profiles). This reflects the realistic diversity of lens models in published works \citep[e.g.,][]{Meneghetti2017,Kelly2023a}. While SN H0pe presents a valuable opportunity to explore lens modeling systematics, as is done in Appendix \ref{app:systematics}, an exhaustive investigation of the full suite of lens modeling decisions is beyond the scope of this work.

\section{Inferring $H_0$ Through Bayesian Analysis} \label{sec:bayes}
We followed a Bayesian approach to constrain $H_0$ similar to the case of SN Refsdal in \cite{Kelly2023a}. For any individual lens model, $H_0$ can be inferred by rescaling the lens model-predicted time delays to match the photometrically and/or spectroscopically measured time delays. There are two measured time delays,  $\Delta t_{a,b}$ and $\Delta t_{c,b}$ for the image a--b and image c--b, respectively. We can infer $H_0$ from both time delays simultaneously, which is  advantageous compared to doing so separately because the time delays may be covariant with one another.
\subsection{From Time Delays Alone}
We begin with a minimal set of observables made up of only the two time delays $\mathcal{O}: \{\Delta t_{a,b}, \Delta t_{c,b}\}$, however this may be expanded to include other observables covariant with the time delays. For simplicity, we first consider only the photometric results. The goal is to determine the probability of $H_0$ given the observed light curve (LC) $P(H_0|LC)$, and marginalize over each lens model $M_l$. The light curve yields measurements of each observable in $\mathcal{O}$, while each lens model $M_l$ makes predictions for the same observables assuming the fiducial $H_0$. For each model, the fiducial $H_0$ is rescaled such that the model predictions best match the measurements from the light curve. Using Bayes' theorem, we can write the probability for the value of $H_0$ from a single lens model as:
\begin{multline} \label{eq:like}
    p_{m}(H_0|{\rm LC}) \propto P(H_0)P(M_l) \int P(\mathcal{O}|M_l ; H_0) \\
    \times P(\mathcal{O}| {\rm LC})d\mathcal{O}_1 ... d\mathcal{O}_n\quad,
\end{multline}
where $n$ is the number of observables. Assuming the systematics of the lens models are not fully independent from one another, the overall likelihood is a weighted sum of the individual model likelihoods:
\begin{multline} \label{eq:like_sum}
    P(H_0|\rm{LC}) = P(H_0) \sum^{N}_{1} \int P(\mathcal{O}|{\it M_l} ; H_0) \\
    \times P(\mathcal{O}| {\rm LC})d\mathcal{O}_1 ... d\mathcal{O}_n\quad,
\end{multline}
where $N$ is the number of lens models. This formalism can be understood as enforcing two priors: 1) following Eq.~\ref{eq:like} each individual lens model will prefer values of $H_0$ that allow the lens model to best reproduce $\mathcal{O}$, and 2) following Eq.~\ref{eq:like_sum} the lens models which best reproduce the observables will have the greatest weight in the sum and hence hold the most influence over the $H_0$ inferred across all lens models. For a set of observables containing only the two time delays, this implies that 1) each lens model will favor values of $H_0$ for which the model-predicted time delays (rescaled following Eq.~\ref{eq:rescale}) best match the two measured time delays and 2) the lens models which best match the ratio of the two time delays are the models which effectively determine the inferred $H_0$.

\subsection{Including Absolute Magnifications}
Constraints on any observables which are correlated with the time delays effectively translate into constraints on the time delays themselves. The SN absolute magnifications are found to correlate with the time delays in the lens models, and, by including them into $\mathcal{O}$, it is possible to infer $H_0$ to greater accuracy and precision than time delays alone. The addition of the image a, b, and c absolute magnifications ($\mu$) increases the total number of observables to five:
\begin{align} \label{eq:obs}
    \mathcal{O}: \{\Delta t_{a,b}, \Delta t_{c,b}, \mu_{a}, \mu_{b}, \mu_{c}\}\quad.
\end{align}

\begin{figure*}[t!]
\includegraphics[scale=0.36]{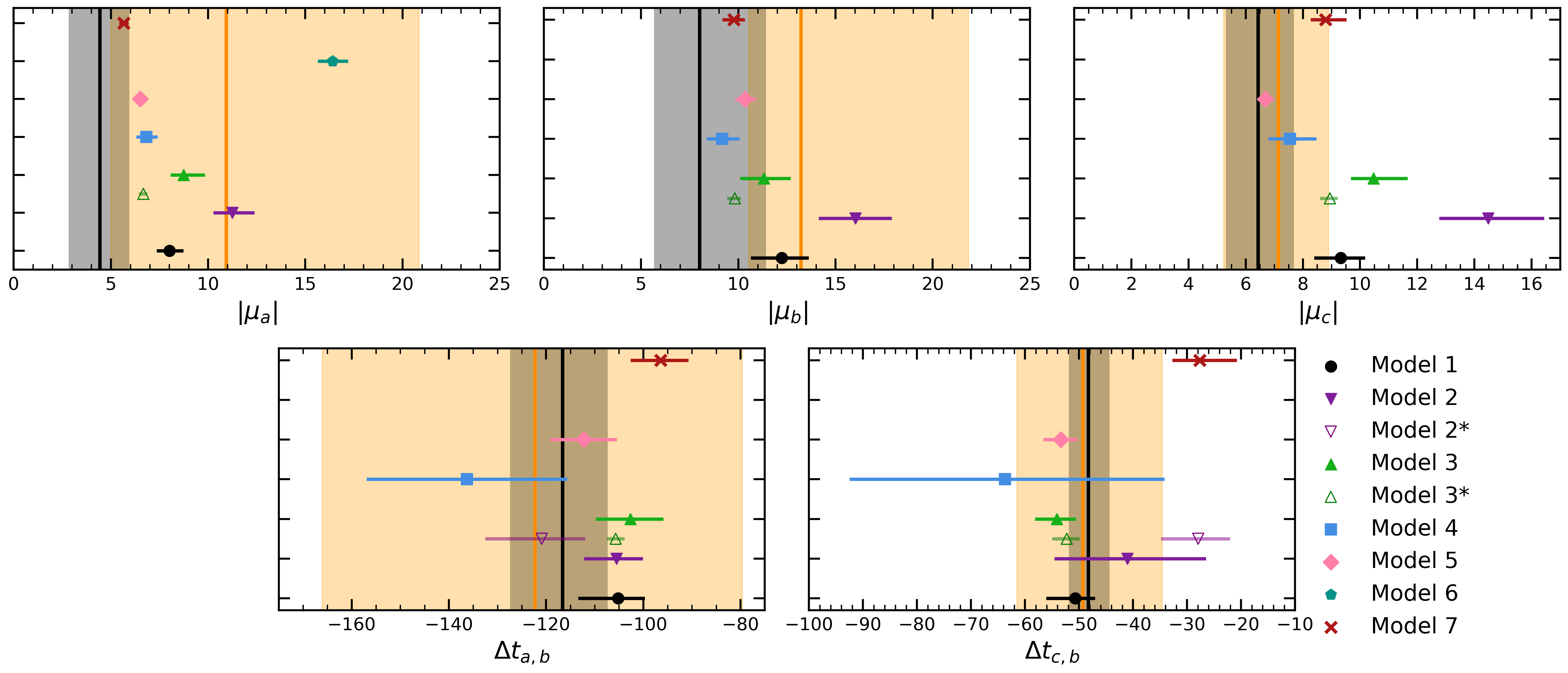}
\caption{The absolute magnifications and relative time delays predicted by each of the seven contributing lens models (markers) and the measured quantities from photometric light-curve fitting (vertical black lines and gray swaths represent the median and 68\% confidence interval respectively) and spectroscopic age-dating (orange lines and swaths). The x-axis limits are set for clarity, which omits model 6 for some quantities due to its inconsistency with both the other lens models and observations. The empty purple and green markers of `Model~2*' and `Model~3*' respectively represent the blinded results from Models 2 and 3 which contained an errors which were amended after unblinding (see \S~\ref{sec:blind}). 
The lens model-predicted time delays assume a fiducial $H_0 = 70~{\rm km}~{\rm s}^{-1}~{\rm Mpc}^{-1}$. 
Following Eq.~\ref{eq:rescale}, the time delays are inversely proportional to $H_0$, and rescaled for a range of values of $H_0$ to infer the total probability distribution.}
\label{fig:lens_res}
\end{figure*}

The measurements for these observables are shown in Fig.~\ref{fig:lens_res} and Table~\ref{tab_models}. Once again following the formalism of Eq.~\ref{eq:like} and Eq.~\ref{eq:like_sum}, the lens model time delay--magnification covariance implies that inducting magnification into $\mathcal{O}$ affects the $H_0$ inferred by each individual lens model, as well as the $H_0$ inferred across all lens models. 
Because the absolute magnification was unavailable for the Type~II SN Refsdal, this marks the first case where it can be meaningfully leveraged for time-delay cosmography. We refer to Sections \ref{sec_ia} and \ref{sec:cov} for further discussion.

\subsection{Including Spectroscopic Constraints}
The spectroscopic data set yields measurements of all five observables independently from the photometric data set. Though the spectroscopic measurements are somewhat less precise,
it is possible to constrain the observables to even greater precision through a joint-fitting approach. Assuming the spectroscopic and photometric constraints are fully independent, the likelihood of Eq.~\ref{eq:like_sum} can be modified to include the spectroscopic measurements, such that we determine the probability of $H_0$ given both the observed light curve (LC), and the observed spectroscopic evolution (SP):

\begin{multline} \label{eq:like_sum_spec}
    P(H_0|\rm{LC,SP}) = P(H_0) \sum^{N}_{1} \int P(\mathcal{O}|{\it M_l} ; H_0) P(\mathcal{O}| {\rm LC}) \\
    \times P(\mathcal{O}| {\rm Sp})d\mathcal{O}_1 ... d\mathcal{O}_n
\end{multline}
Here $P(\mathcal{O}| {\rm LC})$ and $P(\mathcal{O}| {\rm Sp})$ refer to the posteriors for the photometric and spectroscopic constraints respectively.

Altogether, we employ three approaches to the inference of $H_0$ following this formalism. First, we infer $H_{0,{\rm TD-only}}$, which uses only the photometric constraints following Eq.~\ref{eq:like_sum}, and uses only the two time delays in the set of observables, $\mathcal{O}: \{\Delta t_{a,b}, \Delta t_{c,b}\}$. For the second approach, we similarly follow Eq.~\ref{eq:like_sum} with only photometric constraints, but include absolute magnifications, following the full set of observables given in Eq.~\ref{eq:obs} to infer $H_{0,{\rm phot-only}}$. Finally in the third approach, we follow Eq.~\ref{eq:like_sum_spec} to infer $H_{0,{\rm phot+spec}}$, which includes all available constraints from photometry and spectroscopy. We note that $H_{0,{\rm TD-only}}$ inference provided identical results when spectroscopic constraints are added, and hence we do not include it as a separate approach. Across all approaches, we assign a top-hat prior for the Hubble constant: $H_0 \in (30,130)~{\rm km}~{\rm s}^{-1}~{\rm Mpc}^{-1}$. The parametrization of $ P(\mathcal{O}|M_l ; H_0)$, $P(\mathcal{O}| {\rm LC})$, and $P(\mathcal{O}| {\rm SP})$ from the lens model, light-curve fitting, and spectroscopic age-dating posteriors respectively, as well as the integration of the likelihood, are described in Appendix~\ref{app:prob}.

\begin{deluxetable*}{ccccccc}
\tabletypesize{\footnotesize}
\tablecaption{Magnifications and time delays from cluster lens models, photometry, and spectroscopy}
\label{tab_models}
\tablecolumns{7}
\tablehead{
\colhead{\bf\#} &   \colhead{{\bf Lens Model}} & \colhead{\bf $\Delta t_{a,b}$} & \colhead{\bf $\Delta t_{c,b}$} & \colhead{\bf $|\mu_{a}|$} & \colhead{\bf $|\mu_b|$} & \colhead{\bf $|\mu_c|$}} 
\startdata
1 & \texttt{GLAFIC} \citep{Oguri2010,Oguri2021} & $-105.18^{+5.16}_{-7.89}$ & $-50.71^{+3.37}_{-5.09}$ & $8.02^{+0.64}_{-0.57}$ & $12.23^{+1.30}_{-1.51}$ & $9.32^{+0.80}_{-0.87}$ \\
2 & \texttt{Zitrin-analytic}$^b$ \citep{Zitrin2015} & $-105.52^{+5.09}_{-6.32}$ & $-41.06^{+14.27}_{-13.19}$ & $11.25^{+1.05}_{-0.90}$ & $16.03^{+1.77}_{-1.81}$ & $14.48^{+1.91}_{-1.65}$ \\
3$^{c}$ & \texttt{LENSTOOL} \citep{Kneib2011b} & $-102.68^{+6.47}_{-6.80}$ & $-54.09^{+3.18}_{-3.76}$ & $8.75^{+1.01}_{-0.60}$ & $11.32^{+1.29}_{-1.14}$ & $10.47^{+1.14}_{-0.74}$ \\
4 & \texttt{MARS} \citep{Cha2022} & $-136.27^{+20.23}_{-20.32}$ & $-63.72^{+29.24}_{-28.46}$ & $6.82^{+0.50}_{-0.44}$ & $9.17^{+0.80}_{-0.72}$ & $7.55^{+0.86}_{-0.70}$ \\
5 & \cite{Chen2020} & $-112.25^{+6.43}_{-6.57}$ & $-53.35^{+2.71}_{-2.99}$ & $6.52^{+0.24}_{-0.22}$ & $10.35^{+0.46}_{-0.41}$ & $6.68^{+0.24}_{-0.22}$ \\
6 & \texttt{WSLAP+} \citep{Diego2005} & $-273.35^{+95.30}_{-95.30}$ & $342.75^{+92.54}_{-92.54}$ & $16.38^{+0.71}_{-0.71}$ & $47.54^{+3.23}_{-3.23}$ & $33.35^{+0.74}_{-0.74}$ \\
7 & \texttt{Zitrin-LTM} \citep{Zitrin2009} & $-96.45^{+5.41}_{-5.85}$ & $-27.64^{+6.52}_{-4.80}$ & $5.66^{+0.15}_{-0.14}$ & $9.77^{+0.47}_{-0.51}$ & $8.80^{+0.67}_{-0.46}$ \\
\hline
Photometry & \cite{Pierel2023H0pe}$^{a}$ & $-116.60^{+10.83}_{-9.29}$  & $-48.30^{+3.58}_{-3.97}$ & $4.43^{+1.52}_{-1.60}$  & $8.00^{+3.42}_{-2.34}$  & $6.43^{+1.25}_{-1.13}$ \\
Spectroscopy & \cite{Chen2023H0pe}$^{a}$ & $-122^{+43.7}_{-43.8}$  & $-49.3^{+12.2}_{-14.7}$ & $10.93^{+8.63}_{-5.16}$  & $13.22^{+7.49}_{-2.33}$  & $7.14^{+1.53}_{-1.66}$
\enddata
\tablecomments{{\bf Top:} The seven contributing cluster lens models, their primary references, and predicted time delays (in days) and magnifications given as the 68\% confidence interval. {\bf Bottom:} Measurements of the corresponding quantities from photometry and spectroscopy.}
\tablenotetext{a}{Values cited in this table assume Gaussian process regression interpolation for inference of the unlensed SN apparent magnitude.}
\tablenotetext{b}{Time delay values for this model were corrected for an error following unblinding. The error was in the TD measurement and did not require a modification of the underlying blind model. The initial blinded values were $\Delta t_{a,b}=-120.89^{+8.61}_{-11.31}$~days and $\Delta t_{c,b}=-27.94^{+5.56}_{-6.60}$~days. Details on the error and its correction are given in \S\ref{sec:blind}.}
\tablenotetext{c}{This model was updated following unblinding. The blinded model values were $\Delta t_{a,b}=-105.68^{+1.51}_{-1.48}$~days, $\Delta t_{c,b}=-52.27^{+2.06}_{-2.44}$~days, $|\mu_{a}|=6.67^{+0.11}_{-0.14}$, $|\mu_{b}|=9.82^{+0.23}_{-0.31}$, and $|\mu_{c}|=8.94^{+0.21}_{-0.29}$.  Details on the blinded model and its correction are given in \S\ref{sec:blind} and the model 3 section of Appendix \ref{app:lens_models}. }
\end{deluxetable*}

\section{Blinded Analysis} \label{sec:blind}
\subsection{Blinding Protocols}
Blinding protocols were installed to reduce the impact of human bias. They established the information that could shared between subgroups of our team, the live unblinding process, and the rules for which changes were allowable post-unblinding. 
The values of the time delays and the implied magnifications from both light curve analysis and spectroscopic analysis were not revealed to one another, nor to any of the lens modeling teams. The modelers assembled regularly 
to agree upon the set lensing constraints, such as the catalogs of lensed image positions, spectroscopic redshifts, and photometry of the cluster galaxies as stated explicitly in \citet{Frye2023b}, and then constructed the individual lens models independently. All lens models submitted prior to unblinding were required to be included in the inference. More specifically, for the blinded inference of $H_0$, the relative time delays and absolute magnifications from photometry and spectroscopy were blinded via a median subtraction of the probability distribution function (PDF) of each measurement. 

One exception is that the spectroscopic time delay and magnification measurements were unblinded to the lens model 5 predictions. This is because the spectroscopic measurements and the lens model construction were performed by the same team, however the lens model was completed and locked prior to obtaining the spectroscopic results, and hence was blinded from the spectroscopic time delays and the magnifications.

The photometric and spectroscopic analysis timelines were different. To accommodate the staggered timelines, the initial live unblinding was performed only for $H_{0,{\rm phot-only}}$, which the spectroscopic analysis remained blinded to until the $H_{0,{\rm phot+spec}}$ value was unblinded at a later date. Post-unblinding changes were only allowed to address errors.
We defined an error to be a genuine mistake for which there is an objectively correct solution rather than a subjective choice. Following the $H_{0,{\rm phot-only}}$ unblinding, six errors were identified by this definition and corrected. Amending these errors ultimately resulted in a $\sim 1\sigma$ decrease in $H_{0,{\rm phot-only}}$. While the resolution of these errors impacts the inference of $H_0$, the  methodology was unchanged, and the details of each error and its impact on both the observables and the inferred $H_0$ are listed below.

\subsection{Errors Corrected Post-Unblinding}
Here we discuss the initial unblinded value of $H_0$, the seven errors found subsequent to the unblinding, and how amending the errors affected the inference. These errors result from genuine mistakes which had objective fixes. The following events are stated in chronological order beginning with the unblinding event. The initial unblinding was conducted using only photometric constraints, and all errors were found while blinded to the spectroscopic results. Hence, all values given pertain to $H_{0,{\rm phot-only}}$.

The inference of $H_{0,{\rm phot-only}}$ was unblinded live by revealing the median values of the distribution of measured magnifications and time delays from light-curve fitting, which yielded $H_{0,{\rm phot-only}} = 81.3^{+15.9}_{-8.5}$~km~s$^{-1}$~Mpc$^{-1}$.

Post-unblinding, the first error was a miscommunication between the light-curve fitting team and the team performing the $H_0$ inference that resulted in the exclusion of both the ${\sim}0.1$~mag intrinsic scatter of SN~Ia magnitudes and the scatter due to millilensing from the inference. Applying these effects broadened the measured magnification uncertainties by ${\sim}25\%$, which resulted in an updated value of $H_{0,{\rm phot-only}} = 79.6^{+15.0}_{-8.7}$~km~s$^{-1}$~Mpc$^{-1}$.

Second, the light-curve fitting team found an error in the simulations which estimate the systematic uncertainty due to microlensing. The uncertainties were based on the bias and standard deviation for distributions of measured magnifications relative to the simulated magnifications. However, the measured magnifications were recorded as simply the ratio of measured amplitudes, which ignores the additional uncertainty from other light curve parameters. This was updated to match the final method used to measure the true magnifications \citep[see][for details]{Pierel2023H0pe}, which takes the full uncertainty into account. This further increased magnification uncertainties on images $a$ and $b$ by $\sim$10--20\% while image $c$ remained mostly unaffected because of its phase at the second infrared maximum. The same simulations were found to be using a different version of the BayeSN model \citep{Mandel2022} for simulations and fitting, which led to an artificial bias in fitted parameters. Fixing this error decreased the uncertainty of both time delays by $\sim$20\%, altogether resulting in an updated value of $H_{0,{\rm phot-only}} = 76.9^{+14.2}_{-8.2}$~km~s$^{-1}$~Mpc$^{-1}$.

Third, the light-curve fitting team discovered a numerical error in the removal of bias induced by systematics on the magnification and time delays. The median of each retrieved magnification and time delay distribution, created from the simulations discussed above \citep[and see][for more details]{Pierel2023H0pe} was removed from the measured values to account for biases identified by fitting the simulations \citep[following][]{Kelly2023a}. After unblinding, it was discovered that this bias was being removed with a sign error. This shifted the median magnifications by $\lesssim$5\%, the median $\Delta t_{c,b}$ by $\sim$4\%, and $\Delta t_{a,b}$ by $\lesssim$1\%. These resulted in an updated value of $H_{0,{\rm phot-only}} = 75.9^{+15}_{-8.4}$~km~s$^{-1}$~Mpc$^{-1}$.

Fourth, the lens modeling team for model~3 corrected an error in the sampling of the lens model results which had been used for estimation of the posterior distribution in the inference. This was caught serendipitously when the lens modeling team was requested to generate samples for other lens model parameters to gauge the dominant source of lens model uncertainty. The improper sampling resulted in a loss of covariance information between the magnification and time delays and significantly overestimated the time delay uncertainties by a factor of $\sim$5. Re-generating the samples correctly did not require any changes in the lens model itself and did not change the median predictions of the magnifications or time delays. The lens model was given similar weight in the $H_{0,{\rm phot-only}}$ inference both before and after making this adjustment, resulting in an updated value of $H_{0,{\rm phot-only}}=$~\hph.

Fifth, the lens modeling team for model~2 found an error in the calculation of the time delay from the lens model, where the source position sampled was offset by 1 pixel (0\farcs03) in the source plane. This did not require any modification of the underlying lens model, only a change in how time delays were measured from it. As this significantly changed the time delay distribution for this model, both the blinded values and the error-corrected values are shown Table~\ref{tab_models} and Figure~\ref{fig:lens_res}. Both $H_{0,{\rm phot-only}}$ and $H_{0,{\rm phot+spec}}$ were unaffected by this change as a result of negligible weighting both before and after the correction. For $H_{0,{\rm TD-only}}$, however, the model weighting rose from $\sim$0 to $\sim$16\%, changing the inference from $H_{0,{\rm TD-only}}=73.2^{+10.5}_{-6.4}$~km~s$^{-1}$~Mpc$^{-1}$ to $H_{0,{\rm TD-only}}=$~\htd. 

Sixth, during the review process it was determined that lens model 3 included two choices of model parameters which were not sufficiently justified, and ultimately the lens modeling teams mutually agreed these decisions were inappropriate. This specifically referred to the choice of positional uncertainty ($0.03\arcsec$), which was found to be too small compared to prescriptions in the literature \citep[e.g.,][]{Johnson2014,Grillo2016}, and the choice to fix the cluster member reference cutoff radii and velocity dispersions, which were found to bias the results. Updating the model with a more appropriate positional uncertainty of ($0.3\arcsec$) and allowing the cluster member parameters to be freely optimized was found to increase the time delay and magnification fractional uncertainties as much as a factor of 5, and also increased the median magnifications by $\sim 25\%$. The model updates are given in greater detail in Appendix \ref{app:lens_models}. We note that the model was also changed from source-plane to image-plane optimization, as the source-plane model had produced unphysical results at the larger positional uncertainty. Both $H_{0,{\rm TD-only}}$ and $H_{0,{\rm phot+spec}}$ were negligibly affected by this change, however in the $H_{0,{\rm phot-only}}$ inference the model weighting decreased from $28\%$ to $6\%$, resulting in a change from $H_{0,{\rm phot-only}}=74.4^{+13.0}_{-6.0}$~km~s$^{-1}$~Mpc$^{-1}$ to $H_{0,{\rm phot-only}}=$~\hph. 

Seventh, it was found during the review process that the Model 6 magnifications and time delays were measured at the observed SN positions rather than the model-predicted positions, which was inconsistent with the guidelines determined for lens model submission. To ensure a consistent approach across all models, the model 6 results were updated to instead use measurements made at the model-predicted positions. This did not involve any modification of the model itself, nor did this meaningfully affect the $H_0$ inference.
\bigskip

\section{Results} \label{sec:results}
Following the most straightforward approach, we inferred $H_0$ using only the 
photometric time delays, yielding $H_{0,{\rm TD-only}}=$ \htdn~km~s$^{-1}$~Mpc$^{-1}$. In this approach,
 the lens models are weighted based on their ability to reproduce the ratio between the two photometrically measured time delays. This inference is dominated by five models: model 5 (30\%), model 1 (23\%), model 2 (17\%), model 4 (16\%), and model 3 (9\%). 
 We found that including the spectroscopic time delays did not change the result due to their agreement with the photometric time delays while carrying uncertainties that were larger by a factor of $\sim$4.
 Model 6 in particular is truncated by the $130$~km~s$^{-1}$~Mpc$^{-1}$ upper bound of the top-hat $H_0$ prior, resulting in only 3\% of the weighting. Removal of this prior results in a dominant weighting of $\sim 20\%$ for $H_{0,{\rm TD-only}}$ , while the magnification-weighted approaches remain unaffected.

We next included magnification into the suite of observables, which yields $H_{0,{\rm phot-only}}=$~\hphn km~s$^{-1}$ Mpc$^{-1}$ from photometric constraints. Because the lens model-predicted time delays and magnifications are correlated, the magnifications influence both the $H_0$ inferred by an individual lens model, as well as the relative weighting each model receives in the $H_0$ inferred across lens models. The weighting is such that the models which best reproduce the two measured time delays and three measured absolute magnifications simultaneously are given greater contribution to the PDF of $H_{0,{\rm phot-only}}$. The result is primarily influenced by three lens models: model 5 is given 43\% of the weighting, model 4 is given 29\%, and model 1 is given 21\%. No single model dominates the inference, instead the three subdominant contributions individually predict different $H_0$'s which ultimately broaden the resulting PDF.

\begin{figure}[t!]
\includegraphics[scale=0.375]{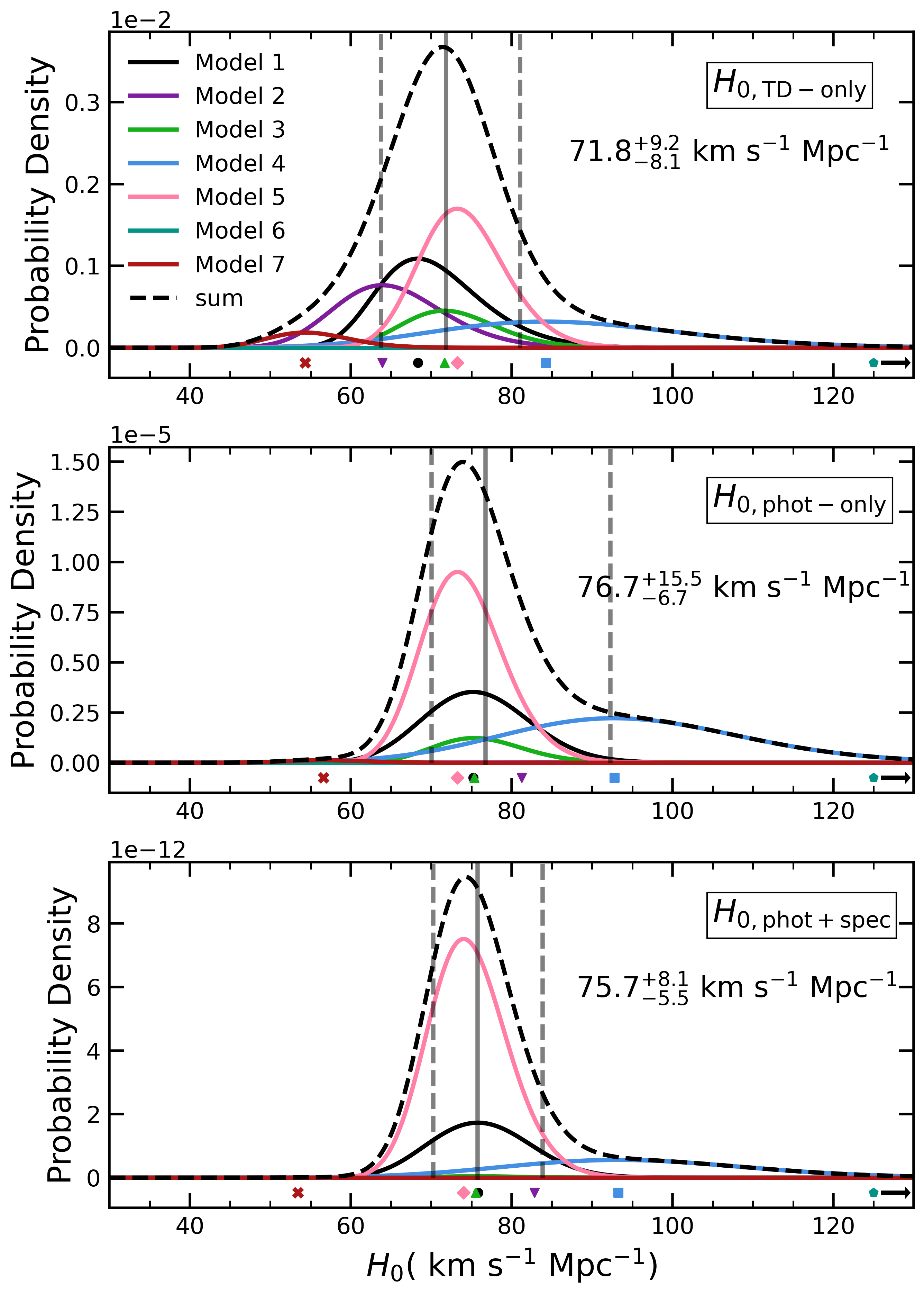}
\caption{Inferred $H_0$ from comparing lens model-predicted time delays and absolute magnifications to those measured from (in order of increasing constraints) only time delays (top), only photometry (middle), and combined photometry and spectroscopy (bottom). The summed PDF (dashed curves) represents the PDF of $H_0$ inferred by the total likelihoods of Eq.~\ref{eq:like_sum} and Eq.~\ref{eq:like_sum_spec}. Individual lens model inferences (solid curves) are shown with normalization equal to their weighting, and the median prediction of each model is plotted as markers below the curves; both the weightings and median $H_0$'s are given in Table~\ref{tab_h0}. The median of model 6 pushes up to the $H_0\leq 130$~km~s$^{-1}$~Mpc$^{-1}$ prior in all cases, and is instead shown with an arrow.}
\label{fig:h0_phot}
\end{figure}

Finally, we obtained the highest precision estimate by making use of all available constraints, inferring $H_{0,{\rm phot+spec}}=$~\hphsp\, from joint photometric-spectroscopic constraints. This inference is dominated by model 5 ($65\%$ of the weighting), with subdominant contributions by model 1 (19\%), and model 4 (14\%).
While the individual lens model predictions are similar to those in $H_{0,{\rm phot-only}}$ (that is, joint time delay constraints are effectively the same as those from photometry alone), the joint magnification constraints favor a single model, resulting in a significantly reduced uncertainty.
For each approach, the $H_0$ predictions of each individual model as well as their assigned weights are shown in Table~\ref{tab_h0}. These results are also illustrated in Figure~\ref{fig:h0_phot}, which shows the PDF of each lens model as well as the total PDF of $H_0$.

\begin{figure*}[h]
\centering
\includegraphics[scale=0.65]{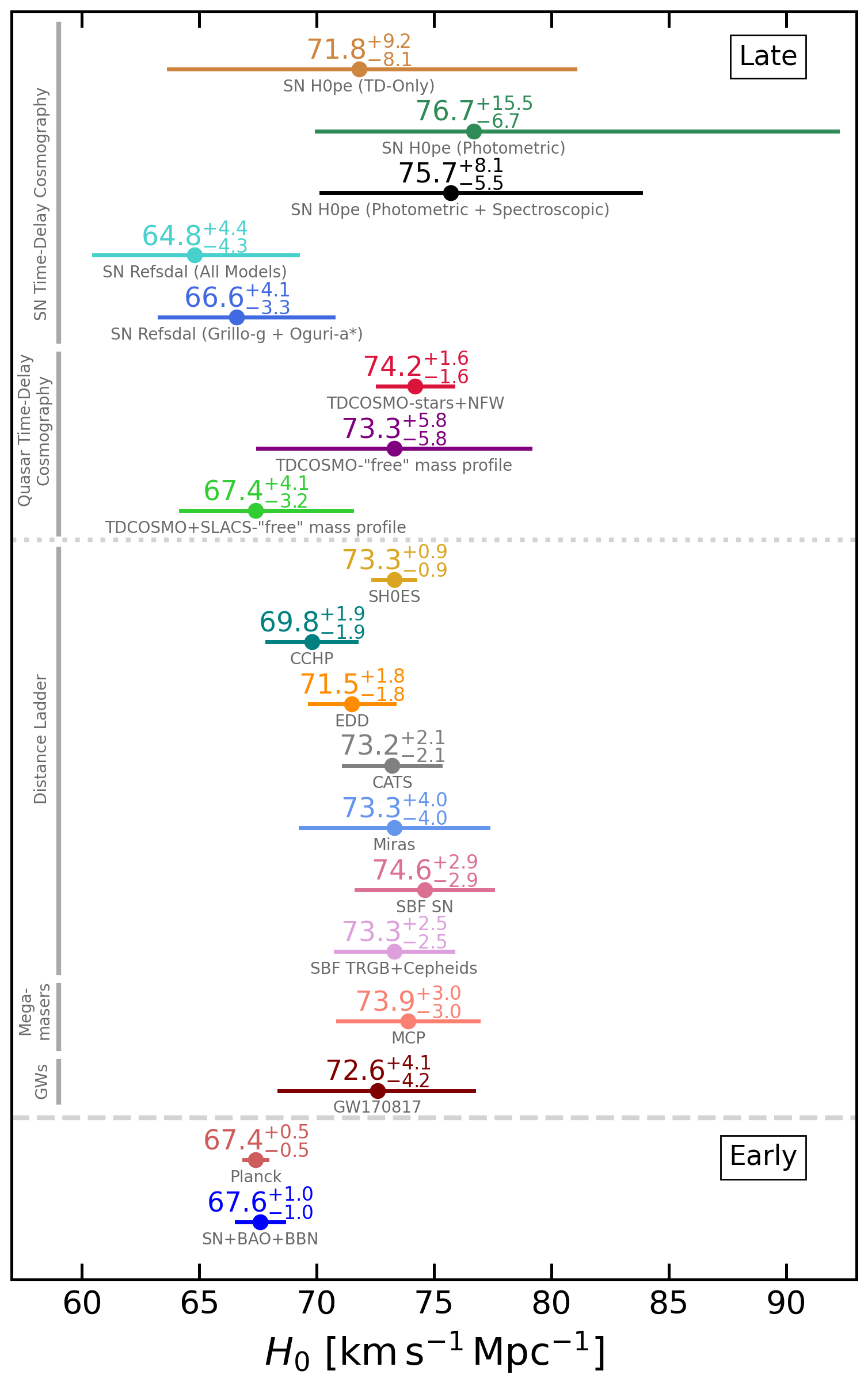}
\caption{$H_0$ measurements from SN H0pe compared to other independent methods, broadly divided into late- and early-time measurements. SN time-delay cosmography measurements include those from SN Refsdal \citep{Kelly2023a} using all contributing lens models and using a subset of two models (Grillo-g and Oguri-a*). Quasar time-delay cosmography measurements include those using fully analytic mass profiles \citep[stars+NFW;][]{Millon2020} and free-form mass profiles \citep[``free" mass profile;][]{Birrer2020} from lenses within the Time-Delay COSMOgraphy (TDCOSMO) collaboration as well as the Sloan Lens ACS (SLACS) sample \citep{Bolton2006}. Distance-ladder measurements include those from SH0ES \citep{Murakami2023}, the CCHP \citep{Freedman2019}, the EDD \citep{Anand22}, CATs \citep{Scolnic23}, Miras \citep{Huang2020}, SBF from SNe \citep{Garnavich2023} and TRGBs+Cepheids \citep{Blakeslee21}. Measurements using  megamasers from the MCP \citep{Pesce20} and using GW170817 as a standard siren \citep{Wang2023} are also included. Early-time measurements include those from CMB anisotropies from the Planck collaboration \citep{Planck2020b} and from large-scale-structure clustering using SNe, baryon acoustic oscillation (BAO), and big-bang nucleosynthesis (BBN) data \citep{Schoneberg2022}. This figure is adapted from \cite{Bonvin2020}.
}
\label{fig:h0_comp}
\end{figure*}

Figure~\ref{fig:h0_comp} compares $H_0$ inferred from SN H0pe to other $H_0$ measurements, including that from SN Refsdal. The combined photometric-spectroscopic inference, $H_{0,{\rm phot+spec}}$, provides the strongest constraints and  agrees with most other late-time measurements, showing less than ${<}1\sigma$ tension with measurements from SH0ES, CCHP, and TDCOSMO \citep{Murakami2023,Freedman2019, Wong2020}. By contrast, SN H0pe's $P(H_0<67.5{\rm~km~s}^{-1}{\rm ~Mpc}^{-1})<7\%$ shows 1.4--1.5$\sigma$ tension with early-time measurements. SN H0pe's $H_0$ is also in 1.3--1.5$\sigma$ tension with the  SN Refsdal result of \cite{Kelly2023a}. However,  further exploration of SN Refsdal's systematics \citep{Liu2024,Grillo2024} 
measured $H_0 = 70.0^{+4.7}_{-4.9}$~\kmsM\ and $H_0 = 65.1^{+3.5}_{-3.4}$~\kmsM. The \cite{Grillo2024} result did not include priors from other cosmological experiments, while this work, \cite{Kelly2023a}, and \cite{Liu2024} all assumed a flat $\Lambda$CDM cosmology.

\begin{deluxetable*}{ccccccc}
\tabletypesize{\footnotesize}
\tablecaption{$H_0$ Inferences}
\label{tab_h0}
\tablecolumns{7}
\tablehead{
\colhead{\bf Lens Model} &  \colhead{\bf $H_{0,{\rm phot+spec}}$} &  \colhead{\bf $H_{0,{\rm phot-only}}$} &  \colhead{\bf $H_{0,{\rm TD-only}}$}  & \colhead{\bf Weight$_{\rm phot+spec}$} & \colhead{\bf Weight$_{\rm phot-only}$} & \colhead{\bf Weight$_{\rm TD-only}$}}
\startdata
1 & $76.05^{+6.41}_{-6.01}$ & $75.55^{+6.71}_{-6.31}$ & $69.74^{+7.11}_{-5.71}$ & $0.19$ & $0.21$ & $0.23$ \\
2 & $83.55^{+6.61}_{-5.91}$ & $82.05^{+7.01}_{-6.11}$ & $64.73^{+7.31}_{-6.31}$ & $0.00$ & $0.00$ & $0.17$ \\
3 & $76.05^{+5.51}_{-5.11}$ & $75.85^{+5.91}_{-5.31}$ & $72.24^{+6.21}_{-5.51}$ & $0.00$ & $0.06$ & $0.09$ \\
4 & $93.96^{+15.02}_{-14.31}$ & $93.66^{+14.91}_{-14.21}$ & $85.96^{+16.02}_{-14.21}$ & $0.15$ & $0.29$ & $0.16$ \\
5 & $74.54^{+5.11}_{-4.70}$ & $73.84^{+5.31}_{-4.80}$ & $73.84^{+5.51}_{-5.01}$ & $0.65$ & $0.43$ & $0.30$ \\
6 & $>130$ & $>130$ & $>130$ & $0.00$ & $0.00$ & $0.00$ \\
7 & $53.92^{+4.90}_{-4.40}$ & $57.03^{+5.01}_{-4.50}$ & $54.82^{+5.31}_{-4.80}$ & $0.00$ & $0.01$ & $0.03$ \\
\hline
Total &  \hphspn\ & \hphn\  & \htdn\  & $1.00$  & $1.00$ & $1.00$
\enddata
\tablecomments{The seven contributing cluster lens models with their individual predictions for the three $H_0$ (km~s$^{-1}$~Mpc$^{-1}$) inferences. Also listed are the relative weightings for each model in the likelihood across the three inferences. We also present the results for the total inference resulting from Eq.~\ref{eq:like_sum} and Eq.~\ref{eq:like_sum_spec}.}
\end{deluxetable*}

\begin{figure}[ht!]
\includegraphics[scale=0.45]{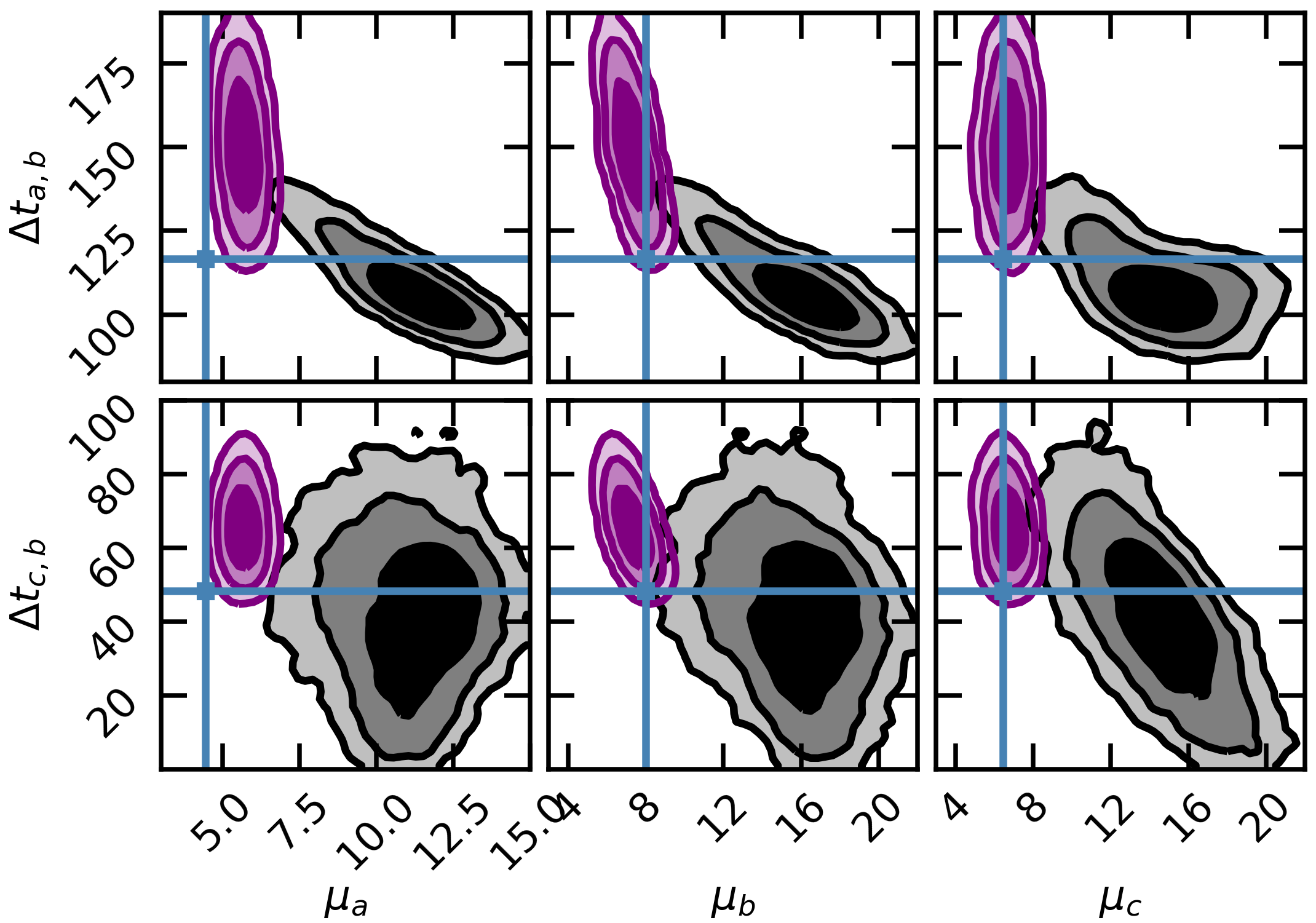}
\caption{2D posterior distributions relating time delays and absolute magnifications predicted by the blinded lens model 2 used in this study (black). The contours represent the 68, 95, and 99\% confidence intervals, and the measured values from light-curve fitting are overplotted (blue lines). As seen in the black contours, the blinded lens model-predicted magnifications and time delays are inversely correlated.
We also plot posteriors for a lens model constructed after unblinding which uses an identical approach to model 2, with the exception that the magnifications measured from light-curve fitting are now applied as constraints in modeling (referred to as an `unblinded' version of the model; purple contours). Including magnification constraints causes the predicted time delays to become larger, following the expectation from the covariance of the blinded model.}
\label{fig:lens_corner}
\end{figure}

\begin{figure}[h]
\includegraphics[scale=0.375]{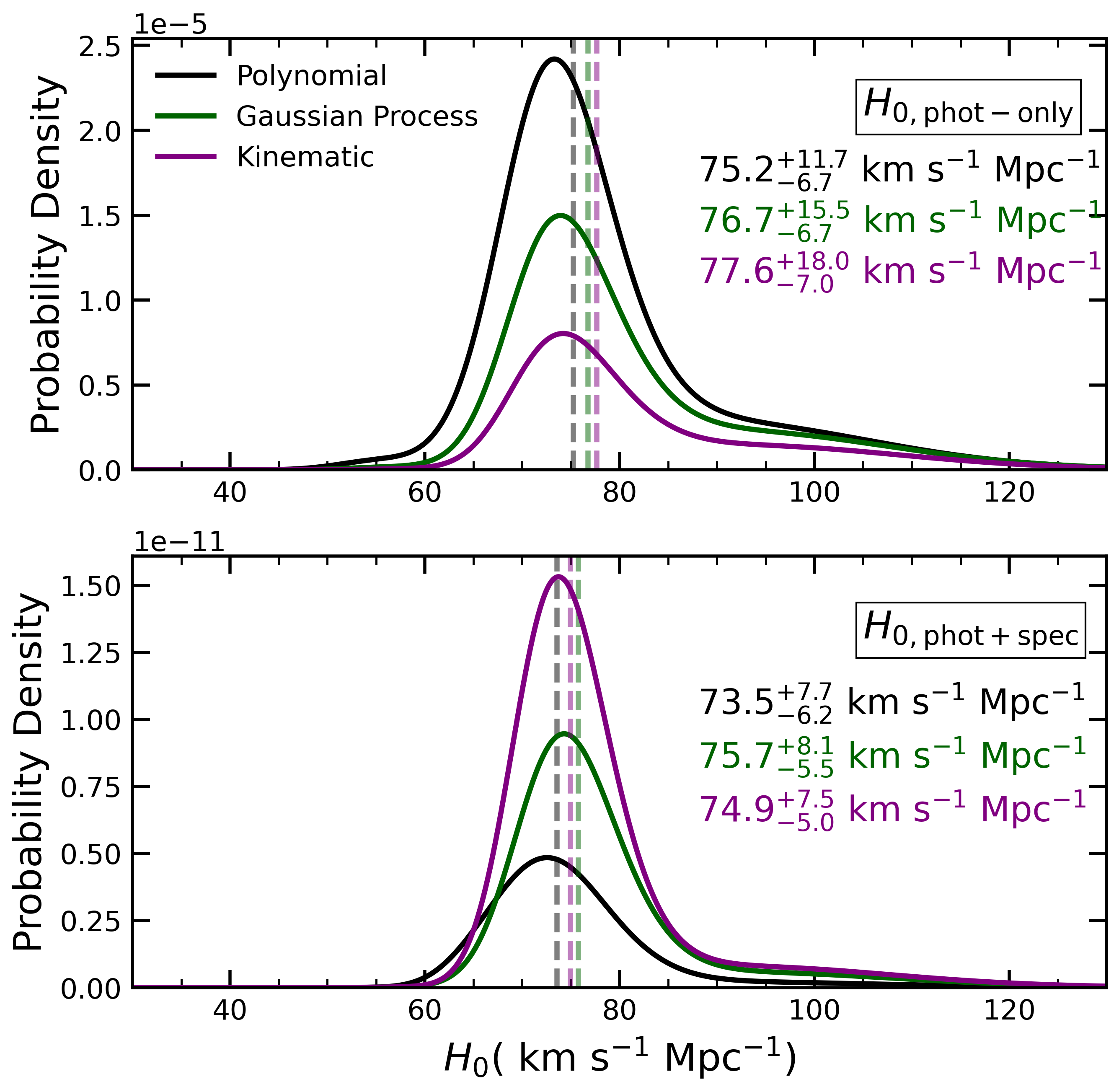}
\caption{The effect of different interpolation methods for estimating the unlensed SNe\,Ia apparent magnitude at $z=1.783$ on $H_0$. We compare the inferred $H_0$ under three interpolation approaches: polynomial, Gaussian process, and kinematic expansion. The interpolation method affects the inferred magnification of each SN image, which in turn influences the $H_0$ inference. The impact is $\sim 3\%$ for both photometric-only constraints and joint photometric-spectroscopic constraints, well within the $1\sigma$ uncertainties.}
\label{fig:interp_h0}
\end{figure}

\section{Discussion} \label{sec:disc}
\subsection{Time-Delay Cosmography With a SN~Ia} \label{sec_ia}
SNe Ia as standard candles offer unique advantages for time-delay cosmography compared to SNe~II such as ranking the lens models via absolute magnification, constraining the impact of mass-sheet degeneracy, and providing an alternative and independent method to measure the time delays and magnifications via spectroscopy.
As a start, we assess the impacts of SN type by repeating the inference of $H_0$, this time by modifying the set of observables to include only constraints available to SNe~II: the photometric measurements of the time delays and the magnification ratios.
This exercise, which makes use of magnification ratios, returns $H_{\rm 0, Type~II} = 73.0^{+11.3}_{-7.8}$~km~s$^{-1}$~Mpc$^{-1}$, providing no meaningful improvement over the inference from time delays alone, $H_{\rm 0,TD-only}$.
This test demonstrates that all the lens models which satisfy the time delay ratios also satisfy the measured magnification ratios to within $1\sigma$. While magnification ratios do not give additional footholds to
break the degeneracy between the lens models and $H_0$,
 absolute magnifications do provide the necessary leverage to rank the lens models. It can be seen that the photometric magnification measurements alone already tighten the overall constraints, downweighting two models which initially satisfied the magnification ratios. On combining the photometric measurements with the spectroscopic measurements, the best constraints of all are obtained, thereby favoring a single lens model (Figure~\ref{fig:h0_phot}). This clearly demonstrates how the magnification helps break this degeneracy and improves the precision of the $H_0$ inference.

\subsection{Lens Model Covariance} \label{sec:cov}
The contributing lens models exhibit varying degrees of covariance between the predicted absolute magnifications and the time delays. The likelihoods of Eq.~\ref{eq:like_sum} and Eq.~\ref{eq:like_sum_spec} depend on both the time delays and magnifications, implying this covariance affects both the $H_0$ inference made by each individual lens model as well as their relative weightings. The lens model-predicted absolute magnifications and time delays are inversely correlated, meaning lower magnifications are associated with larger time delays (2D posteriors in Fig.~\ref{fig:lens_corner}) and therefore larger values of $H_0$.
Among parametric models (1, 2, and 3), this correlation extends between the models, manifesting as an inverse correlation between $H_{\rm 0, TD-only}$ and the predicted magnification of each model. 
\cite{Liu2024} found the same correlation 
for SN Refsdal. The semi- and non-parametric models (4, 5, 6, 7), however, do not follow this correlation across models, affirming they likely probe a different parameter space than parametric models and one that cannot easily be re-created through variations in the parameterization (e.g., the slope of the radial density profile).

Across the full set of models, the parametric models tended to exhibit a stronger correlation between magnifications and time delays than the non-parametric models, perhaps owing to the inflexibility of fully-parametrized mass distributions enforcing a stricter allowable parameter space. We also found that using magnification as a constraint in the lens modeling produces results consistent with the expectations of the covariance (see purple contours of Fig.~\ref{fig:lens_corner}). The specifics of these `unblinded' models are detailed in Appendix~\ref{app:unblinded}.

\subsection{The Mass-Sheet and Related Degeneracies} \label{sec:msd}
Time-delay cosmography can be 
limited by model-independent degeneracies, such as
the mass-sheet and by extension in mass-slope degeneracies, the latter of which is present in  simply-parameterized mass models. This implies that
the time delays and magnifications are sensitive to the addition of constant sheets of surface mass density or variations in the slope of the radial surface-density profile, both of which may still satisfy the lensing constraints
\citep{Kochanek2002,Schneider2013}. 
While these degeneracies tend to plague galaxy--galaxy lenses, in galaxy-cluster lenses they can be broken by the availability of 
two or more multiple-image systems at different spectroscopically confirmed redshifts \citep{Falco1985, Kawano2006, Schneider2014}. For example, \cite{Grillo2020} and \cite{Grillo2024} examined the case of SN Refsdal in the MACS J1149 cluster field and concluded that the availability of 89 multiple images from 27 distinct multiple-image systems (8 with unique spectroscopic redshifts) 
diminished these degeneracies to insignificant levels. However, \cite{Liu2024} found for a similar set of lensing constraints (but different lens modeling approach) that the mass-slope degeneracy, while 
diminished, could be still further decreased by the incorporation of new multiple-image systems.

G165 presents five systems at three unique spectroscopically confirmed redshifts and an additional sixteen multiple image systems with photometric constraints. These constraints already break these degeneracies, however it is unlikely that they are completely eliminated, as is illustrated by the 
scatter in our seven model-predicted time delays. Furthermore, as pointed out in Section \ref{sec_ia}, many of the lens models satisfy the measured magnification ratios yet do not agree with the measured magnification, an indication of the mass-sheet degeneracy at play. This further emphasizes the importance of obtaining a magnification constraint, which has been widely-advocated for resolving this degeneracy in galaxy--galaxy and galaxy-cluster lenses alike \citep[e.g.,][]{Oguri2003,Foxley2018,Birrer2022,Liu2024}.

As discussed in \S\ref{sec_ia}, a magnification constraint is effective at breaking the degeneracy among lens models, reflecting its leverage over the mass-sheet degeneracy. It is interesting to ask to what degree the measured magnification precision can constrain such a mass-sheet. The mass-sheet degeneracy, or more generally the mass-sheet transformation  allows for a mapping of the surface mass density $\kappa^\prime = \lambda\kappa + (1-\lambda)$ which induces a change in the true source position $\beta^\prime = \lambda \beta$ while leaving the apparent source position invariant. The second term in the convergence transformation, $(1-\lambda)$, makes up a constant sheet of surface mass density which gives the MST its name; we refer to this term as $\kappa_\lambda$.  This directly scales the magnifications $\mu^\prime = \mu/\lambda^2$ and time delays $\Delta t^\prime = \lambda\Delta t$, however it is clear the ratios remain invariant. From photometry and spectroscopy, the magnifications are generally measured to an average fractional precision $\sim 25\%$. This implies that a single magnification measurement can rule out $|\Delta \lambda| \gtrsim 0.1$ or directly $|\kappa_\lambda| \gtrsim 0.1$. When combining all three multiple images, this holds the potential to be much stronger. For example we evaluate lens model~5, which gives the best agreement with measured magnifications, and in post-processing apply a range of $\lambda \in [0.8,1.2]$ to its predicted magnifications. We find that only a very small lambda, $\Delta\lambda \sim 0.05$, is required to reduce the model weighting by a factor of two, implying only a very small allowance of $|\kappa_\lambda| \lesssim 0.05$.

We emphasize this test is not comprehensive: an external convergence need not present itself as a constant sheet of mass \citep{Khadka2024}, and it alone does not quantify the extent to which the availability of many multiple image systems breaks this degeneracy. In the case of an individual lens model, the degeneracy could be quantified explicitly by incorporating free parameters for the addition (or subtraction) of sheets of constant surface mass density or for varying the slope of the surface mass density profile during optimization \citep[e.g.,][]{Grillo2020,Grillo2024}. This could capture degeneracies between an external convergence and the lens model parameters, which are not addressed in our test. For example, such an external convergence could result in a fundamentally different lens model with a distinct $\kappa$ distribution (e.g., the isodensity contours are not preserved). Similarly, testing a wide variety of mass profile parametrizations, such as in \cite{Liu2024}, could illustrate the degree of any persisting mass-slope degeneracy. Incorporating magnification constraints into these tests gives additional opportunities to explore lensing degeneracies at the cluster lensing scale, but is outside the scope of this work.

\cite{Kawano2006} found time delay ratios can provide an alternate route to breaking the mass-slope degeneracy. This constraint is rarely highlighted in the literature, but its power is apparent in the inference of $H_{\rm 0, TD-only}$ in this study, where the time delay ratio alone determines the model weighting. This weighting provides a $\sim$12\% fractional uncertainty in the $H_{\rm 0, TD-only}$ measurement, but removal of this weighting scheme (i.e., the PDFs of the models are rescaled to the same normalization) results in uncertainties which are larger by more than a factor of two. This demonstrates the utility of multiple image systems for which multiple precision time delays are available, as these hold an under-recognized ability to break lensing degeneracies.

\subsection{SN~Ia Apparent Magnitude Interpolation} \label{sec:interp}
To ensure the magnifications for SN H0pe are measured independently of cosmology, both the spectroscopic and photometric methods compared the peak apparent magnitude of each SN image to the distribution of field SNe\,Ia.
Due to the relatively small number of spectroscopically confirmed SNe\,Ia at $z>1$ \citep{Scolnic2018}, the field SN~Ia peak magnitudes had to be fit with a smooth model to enable an interpolated inference at $z=1.783$. Before the un-blinding step in this analysis a choice was made to perform this fit with a Gaussian process, but after the un-blinding we explored two additional options for the fitting to understand the impact for our measurement. We fit the field SN~Ia peak magnitudes with a second-order polynomial in $log(z)$-space (allowing all coefficients to vary), as well as with a second-order cosmology-independent kinematic expansion \citep[see][Equation 4]{Riess2022}.

We find that the range predicted by these three methods is fairly large, $\sim0.3$mag, but that the Gaussian process result is situated directly in the center of the polynomial and kinematic expansion fits \citep[see Figure~6 of ][]{Pierel2023H0pe}. We cannot change this method after un-blinding, as there is no evidence to suggest any of the three approaches are in error. 
Nevertheless we evaluate the impact of the interpolation method on the $H_0$ inference, as this value directly translates to the magnification measurements and therefore affects the $H_0$ prediction. The inferences across the three interpolation methods are shown in Figure~\ref{fig:interp_h0}; the median $H_0$ varies at most 3\% in both the photometric and the joint photometric-spectroscopic inferences. As more SNe\,Ia at $z>1.5$ are spectroscopically confirmed, the three methods are expected to converge while still remaining independent of cosmology.

\subsection{Error Budget} \label{sec:disc_err}
The lens model predictions and the photometric plus spectroscopic measurements both contribute significantly to the error budget, resulting in the $\sim$9\% uncertainty in the joint photometric--spectroscopic $H_0$ inference (Table~\ref{tab_h0} column~2). Lens model~5 dominates the weighting in this inference and by itself gives $\sim$6.5\%  uncertainties for the $H_0$ (see the model 5 row of Table~\ref{tab_h0}). The difference between these uncertainties confirms that disagreements among lens models drive up the uncertainty. This represents the systematic uncertainty of the lens models, which could be mitigated stricter magnification constraints from photometry and spectroscopy or by including new lensing constraints \citep[e.g.,][]{Liu2024}. We found the removal of image systems which did not have spectroscopic redshifts worsened the agreement between the lens-model predicted magnifications and observations, demonstrating that even photometrically-selected image systems can provide impactful constraints on the lens models. Thus the addition of new photometric systems, or spectroscopic confirmation of existing systems may both serve to increase lens model precision.

The photometric  time delays each carry $\sim$8\% uncertainty, but the inclusion of two time delays is more constraining than a single one of similar precision. To estimate the error budget, the two time delays can be approximated as a single effective time delay with  uncertainty ${\sim} 8\% /\sqrt{2} = 5.6\%$ (following error propagation through Eq.~\ref{eq:rescale}). Indeed,  an $H_0$ inference assuming zero uncertainty from the lens models (i.e., taking the time delays of lens model~5 with zero error bars) gives an uncertainty on $H_0$ of 5--6\%, consistent with the expectation from the single effective time delay. Taking this approximation, we can straightforwardly decompose the $\sim$9\% $H_0$ uncertainty into two-fifths contribution from photometry and spectroscopy and  three-fifths contribution from lens modeling. Furthermore, 
the lens model uncertainty can be decomposed into one-fourth `measurement' uncertainty (originating from a single model) and three-fourths `systematic' uncertainty (the spread between contributing models).

As shown above, ignoring the lens model uncertainty altogether still results in a modest $H_0$ uncertainty, and therefore improvements in photometry or spectroscopy  still matter. Approved follow-up {\it JWST} Cycle 3 imaging (PID: 4744) after SN H0pe has faded will provide a template image for precision difference-image photometry, which is expected to  improve magnification constraints and nearly halve photometric time delay uncertainties. This could result in as small as a 2.8\% effective time delay uncertainty. Given an effective lens model time delay uncertainty 7\% (which approximately accounts for the model~5 uncertainties and smaller contributions of other lens models), this would translate to $\sim$7.5\% uncertainty on $H_0$. Should the improved magnification constraints from photometry prove to entirely favor a single lens model,  this could push the uncertainty to $\lesssim5\%$.

\section{Conclusions} \label{sec:concl}

SN H0pe is only the second opportunity for precision SN time-delay cosmography. We measured the probability distribution of $H_0$ by comparing the photometrically and spectroscopically measured time delays and magnifications to those predicted by seven independently created lens models, inferring $H_0=$~\hphsp\,.
We found the lens model-predicted time delays and magnifications were correlated, such that including magnifications could break degeneracies between the lens models and $H_0$, influencing both the $H_0$ prediction of each individual lens model as well as weighing its contribution to the total inference. We demonstrated this constraint is a powerful tool to select for lens model accuracy, disfavoring lens model predictions that otherwise would have been 
admitted
had the SN been a Type~II. The lens model predictions and combined photometric-spectroscopic measurements both contribute significantly to the error budget, resulting in the $\sim 9\%$ uncertainty in the joint photometric-spectroscopic inference. 
Follow-up epochs of  {\it JWST} NIRCam imaging 
after SN H0pe has faded are approved, and
will improve the precision of the SN photometry, and hence also both the
time delay and magnification estimates. This combined with the additional constraining power provided by two precision time delays may allow for constraints at the $\lesssim5\%$ level, comparable to or exceeding that of SN Refsdal.

SN H0pe is the first cluster-lensed SN where the methodology for the $H_0$ inference could be developed in advance of the lens model construction and photometric and spectroscopic measurements. All lens modelers received identical lensing constraints, providing a novel opportunity to ensure consistency between lens models and test lens systematics. We found it was crucial to establish rules for blinding due to the number of independent teams contributing to the analysis, which included a formalized process to address errors. We distinguished objective errors from subjective choices, and only admitted changes to the former following unblinding. These protocols were successfully implemented for SN H0pe and provide a timely point of reference for growing the catalog of $H_0$ measurements from cluster-lensed SNe.
Upcoming surveys from the Vera C. Rubin and Nancy Grace Roman observatories as well as continuing  {\it JWST} programs will increase the sample of cluster-lensed SNe, which promises to ultimately drive $H_0$ constraints down to percent-level precision.

\section{Acknowledgements}
We would like to highlight the teams who contributed the lens models for this work, which include Masamune Oguri (model 1), Ashish K. Meena, Nick Foo, and Adi Zitrin (models 2 and 7), Patrick S. Kamieneski (model 3), Sangjun Cha and M.~James Jee (model 4), Wenlei Chen (model 5), and Jose M. Diego (model 6). We thank Chris McKee, Aliza Beverage, Isaac Malsky, James Sullivan, and Emma Turtelboom for useful conversations. \textbf{We also thank the anonymous referees for helpful comments which improved the manuscript}. We thank the JWST Project at NASA GSFC and JWST Program at NASA HQ for their monumental effort in ensuring the success of the JWST mission. This work is based on observations made with the NASA/ESA/CSA James Webb Space Telescope. The data were obtained from the Mikulski Archive for Space Telescopes (MAST) at the STScI, which is operated by the Association of Universities for Research in Astronomy, Inc., under NASA contract NAS 5-03127 for JWST. These observations are associated with JWST programs 1176 and 4446. This work is also based on observations made with the NASA/ESA Hubble Space Telescope (HST). The data were obtained from the Barbara A. Mikulski Archive for Space Telescopes (MAST) at the STScI (\textbf{accessible via \dataset[DOI: 10.17909/zk1p-2q51]{https://doi.org/10.17909/zk1p-2q51}}) which is operated by the Association of Universities for Research in Astronomy (AURA) Inc., under NASA contract NAS 5-26555 for HST. This work was supported by JSPS KAKENHI Grant Numbers JP22H01260, JP20H05856, JP22K21349. M.J.J. acknowledges support for the current research from the National Research Foundation (NRF) of Korea under the programs 2022R1A2C1003130 and RS-2023-00219959. AZ acknowledges support by Grant No. 2020750 from the United States-Israel Binational Science Foundation (BSF) and Grant No. 2109066 from the United States National Science Foundation (NSF); by the Ministry of Science \& Technology, Israel; and by the Israel Science Foundation Grant No. 864/23. P.L.K. acknowledges funding from NSF grants AST-1908823 and AST-2308051. AG acknowledges support from the Swedish National Space Agency, Dnr 2023-00226. 

\bibliography{G165_blf}{}
\bibliographystyle{aasjournal}



\begin{appendix}
\section{The Seven Cluster Lens Models}\label{app:lens_models}
All models were constructed using the same set of lensing constraints and were blinded from one another as well as from the photometrically and spectroscopically measured time delays and magnifications. The exception is model~5, for which the modelers knew the spectroscopic results but did not include them in the lens model constraints. Model 7 was constructed prior to the unblinding, but was not present at the unblinding owing to logistical difficulties. The model was later added to the analysis. Table~\ref{tab_lens_info} gives goodness-of-fit statistics and modeling parameters. The lens models all showcase relatively good quality fits as evidenced by the small root-mean-squared (RMS) offsets between modeled and observed multiple images and the general agreement among the lens models' predictions.

\textbf{Model 1} implements the \texttt{GLAFIC}
software \citep{Oguri2010,Oguri2021}, which utilizes a parametric approach to solve for the analytic lens profiles. \texttt{GLAFIC} adopts an adaptive-mesh grid to increase computation efficiency, essentially increasing resolution in regions with larger magnification gradients (e.g., near critical curves). The model used five Navarro--Frenk--White (NFW) profiles to model the dark-matter halos of the cluster. These were chosen to correspond to the 5 bright galaxies which make up the centers of the two merging mass components. The cluster galaxies were modeled with pseudo-Jaffe ellipsoid profiles scaled according to each galaxy's observed F200W flux density. An external shear component was included to improve fitting flexibility.
Higher-order multipole perturbations  $m=3$ and 4 (where $m=2$ represents the external shear) were also included. Free parameters were also included for the redshifts of the 16 photometrically-constrained multiple image systems. The optimization was performed using an image plane $\chi^2$ approximately estimated in the source plane (See  \citealt{Oguri2010} appendix) using an MCMC with $\sim10^4$ steps. This approximate $\chi^2$ has been found to well-correlate with the true image plane $\chi^2$ (Fig.~5 of  \citealt{Oguri2010}).

\textbf{Model 2} was created using a revised version of the parametric approach detailed by \cite{Zitrin2015}. The new version (sometimes referred to as \texttt{Zitrin-Analytic}; see also \citealt{Furtak2023MNRAS.523.4568FLensModel})  is no longer limited to a grid resolution. This approach operates similarly to other parametric lens modeling techniques and has been successfully applied to  {\it JWST} imaging of other clusters \citep[e.g.,][]{Pascale2022b,Meena2023}. The model employs two primary components, each described by an analytic profile. The cluster galaxies were parametrized as double pseudo-isothermal elliptical mass-density distributions \citep[dPIE,][]{Eliasdottir2007} scaled by their luminosities following common relations. The cluster-scale dark matter halo was parametrized with a pseudo-isothermal elliptical mass-density distribution \citep[PIEMD;][]{Kassiola1993}. Constituting a minor component, the masses of four central galaxies were scaled independently to allow further flexibility (($\alpha,\delta$) = (11:27:06.6969,+42:27:50.363), (11:27:20.0475,+42:29:08.292), (11:27:14.4389,+42:28:14.066), (11:27:14.5094,+42:28:07.004)). While the model can incorporate galaxy ellipticities, for the current model all cluster galaxy profiles were assumed to be circular. In a suite of self-simulated mass distributions, both image- and source-plane optimizations accurately recovered the input parameters, fulfilling the expectation that these should perform similarly \citep{Keeton2010}. For G165, the model optimization was performed in the source plane via a Markov Chain Monte Carlo, which was sampled to derive the uncertainties. The MCMC consisted of $\sim 100$ initial chains of $\sim 10^3$ steps, which is then refined by a final chain run for $\sim 10^4$ steps. For more details on the methodology, including scaling relations and minimization, we refer to \citet{Furtak2023MNRAS.523.4568FLensModel}.

\textbf{Model 3} was constructed using \textsc{Lenstool}\footnote{\url{https://projets.lam.fr/projects/lenstool/wiki}}, which applies parametric mass profiles constrained straightforwardly by the observed positions of the multiply-imaged systems. In this case, the only mass profiles employed were the $z\approx0.35$ cluster members identified by \citet{Frye2023b} plus two cluster-scale halo profiles for the dominant components of the merging cluster, all having a PIEMD  profile. Positions and shape parameters (ellipticity and orientation) of the cluster members were held fixed to their F200W values. The mass of each cluster member was scaled in proportion to its F200W flux, following \cite{Limousin2005}. A characteristic $L_\star$ galaxy at the cluster redshift ($m_{\rm F200W} = 17.0$ AB mag) was set to a velocity dispersion $\sigma_0^\star$, a core radius $r_{\rm core}^\star$, and a cutoff radius  $r_{\rm cut}^\star$. The core radius was fixed to $r_{\rm core}^\star = 0.15$~kpc similar to other strong-lensing works \citep[e.g.,][]{Johnson2014}, while the velocity dispersion and cutoff radii were left to be freely optimized by the model. The masses were fixed according to their scaling relative to $L_\star$ with the exception of the BCG and a perturber cluster galaxy near Arc 3 (($\alpha,\delta$) = (11:27:14.304,+42:28:32.32)), whose scalings were optimized independently. For the two cluster halos, their positions, ellipticities, orientations, velocity dispersions (mass), and core radii were left free to vary. The halos' cut radii are poorly constrained and therefore were held fixed at 1~Mpc. The redshifts of the 16 photometrically-constrained multiple image systems were left to be freely optimized by the model. This model was optimized in the image plane with a positional uncertainty $\sigma_{\rm pos} = 0.3\arcsec$, which was chosen to match the RMS of earlier model iterations.This model was optimized using MCMC, and was run with 10 chains of 1000 steps each, which was found to be much longer than the burn-in time of $\lesssim100$ steps. For further model information, we refer to \cite{Kamieneski2024}.

The model detailed above was constructed following unblinding, correcting a number of issues in the initial blinded model. The blinded iteration fixed the reference values for the cluster member scaling relations to a velocity dispersion $\sigma_0^\star = 120~{\rm km~s}^{-1}$ and a cutoff radius  $r_{\rm cut}^\star = 30$~kpc. While these values were commensurate with findings from other works \citep[e.g.,][]{Limousin2008,Mahler2022,Furtak2023MNRAS.523.4568FLensModel}, it was found to bias the model toward lower magnifications. Leaving these to be freely optimized instead found a much different, but reasonable solution $\sigma_0^\star = 210~{\rm km~s}^{-1}$ and $r_{\rm cut}^\star = 8$~kpc.

The blinded iteration of the model was run via source plane optimization with a small positional uncertainty $\sigma_{\rm pos}=0.03\arcsec$. While this was chosen to match the astrometric uncertainty of the observed lensed-image centroids (which were independently assigned to the same image pixel by multiple team members), previous works using parametric models indicated that this likely underestimated the uncertainties of the model \citep[][]{Zitrin2015,Grillo2016}. The source plane approach, which approximated an image plane chi-squared (see Eq.~12 of \citealt{Jullo2009}), was found to also underperform the image plane optimization, producing an RMS nearly a factor of two larger, and suffered from unphyiscal solutions at the larger positional uncertainty. The source-plane model also showed evidence for being generally less robust: for example, the posterior median was found to change with positional uncertainty (which was not seen in the image-plane variant). For these reasons, the corrected model was run in the image plane, which may have implications for the model posteriors (see Appendix \ref{app:spvsip}). The results of the blinded model are noted in Table \ref{tab_models}, and the impact of its correction on the $H_0$ inference is detailed in Section \ref{sec:blind}.

\textbf{Model 4} was constructed by the \texttt{MARS} algorithm \citep{Cha2022}, which is a free-form,  grid-based approach. \texttt{MARS} consists of two terms, one a chi-square minimization of strongly-lensed (SL) multiple images in the source plane and the other a regularization term. The \texttt{MARS} algorithm adopts maximum cross-entropy for the regularization. This approach can suppress fluctuations and generate a quasi-unique solution for a given SL multiple images. For lens modeling, \texttt{MARS} can consider not only SL multiple images but also weakly-lensed (WL) galaxies  \citep{Cha2023}. To account for WL galaxies, \texttt{MARS} utilizes an additional term 
that minimizes differences between observed and predicted reduced shear. For this study,  \texttt{MARS} provided lens models from both SL-galaxies-only and SL+WL datasets to predict magnifications and time delays. The SL-galaxies-only model consists of a $100\times100$ pixel grid covering $90\arcsec\times90\arcsec$ around the two BCGs. The SL+WL model consists of a $400\times400$ pixel grid encompassing the entire $360\arcsec\times360\arcsec$ FoV\null. One of the interesting differences between the SL-only and the SL+WL reconstructed mass distributions is that the SL-only model shows an offset of the mass peak from the northeastern core because of the deficiency of SL multiple images, but the SL+WL mass map exhibits good agreement with the luminosity distributions of the galaxy-cluster members. 

\textbf{Model 5} uses the lens modeling software originally developed for a blind prediction of the SN Refsdal reappearance and time delays \citep{Chen2020}. This approach makes minimal assumptions about the mass distribution at the galaxy-cluster scale, similar to non-parametric methods, but uses a parametric approach at the individual-galaxy scale. This semi-parametric method parameterizes the dark-matter halos of the  cluster galaxies as symmetric analytic Navarro--Frenk-White profiles \citep{Navarro1997} with the same scale radius and assumes the halo mass scales in proportion to stellar flux. The scale radius is set to 3.84~kpc, and the mass-luminosity ratio for $r<30$~kpc, M/L~($r<30$~kpc) = 9.4 ${\rm M}_\odot/{\rm L}_\odot$. The luminosity is measured from {\it JWST} F090W photometry (approximately V/R band rest-frame). The goal of the cluster member halos is to serve as an initial condition of a concentrated mass distribution around the core of each cluster member, as this cannot be well-constrained by observed multiple images due to the cluster-scale lensing. These profiles do not hold after many iterations of the global perturbation function described below, such that the final solution of the cluster-scale mass distribution is only weakly dependent on the choice of scale radius and halo mass.
 For the cluster-scale dark-matter distribution, the model makes no assumptions about symmetry. Instead, the global lensing potential is deformed by a set of curving functions that apply numerical perturbations to the potential. These curving functions are inherently smooth on the cluster scale and are applied to determine a source plane position that minimizes the separation between source-plane positions of lensed multiple images of image families. The model aims to find the best source-plane position for each set of multiple images that can minimize the image-plane offset. Thus, although the perturbations are driven by the source-plane positions, the model is still optimized using image-plane statistics. While the only previous application of this model was to the MACS J1149.5+2223 cluster, its predictions of the reappearance position of SN Refsdal's image SX as well as Refsdal's four relative time delays and magnification ratios are consistent with other well-tested parametric models \citep{Kelly2023a}. This model does not optimize for the redshifts of photometrically constrained multiple image systems, but instead these systems are assigned to their photometric redshifts and are given a factor of 2.5 lower weighting in the model optimization compared to systems with spectroscopic redshifts.

\textbf{Model 6} is a hybrid  model, implementing the \texttt{WSLAP+} \citep{Diego2005} software, where the large-scale component of the lens model is described by a grid of Gaussians, and the small-scale component  traces  the light distribution of member galaxies. The model is optimized by solving a system of two linear equations per lensing constraint (one for the $x$ coordinate and one for the $y$ coordinate) of the form $\theta=\beta-\alpha(\theta,M)$ with $\theta$ the observed position of the lensed galaxy, $\beta$ the unknown position of the galaxy in the source plane, and $\alpha(\theta,M)$ the deflection angle at the position $\theta$ produced by the unknown mass distribution $M$. 

For G165, the distribution of grid points for the smooth component was built recursively with a first solution obtained from a uniform grid that was then used to derive an adaptive grid that increased the spatial resolution in regions with higher mass concentrations. Using the adaptive grid, \texttt{WSLAP+} derived 20 solutions varying the initial guess from the optimization and the redshifts of the systems with photometric redshifts. These models were used to derive predicted magnifications and time delays at the three model-predicted positions of  SN H0pe. The lack of lensing constraints in the east portion of the cluster and  between the two groups resulted in large variations in the predicted magnifications and time delays. In particular, the critical curve at $z=1.78$ exhibits a dual behavior with some models predicting a single critical curve that connects both groups and other models predicting two disjoint critical curves, one around each group. Model 6 is the only lens model which predicts image C of the SN to arrive last, which is inconsistent with measurements from both photometric light curve fitting and spectroscopic age-dating. This results in the model receiving zero weighting across all three $H_0$ inferences, as each make use of time-delay ratio constraints. Across the suite of solutions used to derive the model uncertainties, some solutions do correctly predict the image arrival order, indicating that new lensing constraints may improve the accuracy of the model.

\textbf{Model 7} uses an updated version of the \texttt{Zitrin-LTM} software (\citealt{Zitrin2009}, \citealt{Zitrin2015}; see also \citealt{Broadhurst2005}), which was modified to accommodate time-delay cosmography of galaxy cluster lenses following SN Refsdal \citep[see][]{Zitrin2020}. In this semi-parametric approach, it is assumed that the light traces the mass (LTM), such that observed cluster-member light is assumed to trace the underlying dark matter. The stellar mass of each cluster galaxy is represented analytically as power-law mass surface-density profile with relative weights determined by the galaxy F200W brightness. The profiles are then simultaneously smoothed with a Gaussian kernel to approximate the projected cluster-scale dark matter distribution, which receives a free weighting relative to the parametric galaxy component. An external shear component with free amplitude and position angle is also applied to broadly account for systematics like large scale structure. Finally, additional flexibility is included for individual galaxies of interest, which included two nearby perturber galaxies specified in Sec. \ref{sec:models}, and the three brightest central galaxies. For the bright galaxies, the position angle, ellipticity, and mass-to-light ratio can be fit freely. Five of the image systems were set to their spectroscopic redshifts, and for all other systems the redshifts were left free in the model, using the photometric redshift estimate of \cite{Frye2023b} as an initial guess. The minimization of the model is done in the image plane, using a MCMC with a final chain of 2000 steps from which the posteriors are sampled.

\section{Lens Model Positional Uncertainties} \label{app:pos_unc}
The positional uncertainties of the lens models play a determining role in the time delay and magnification errors of the parametric lens models of this sample, namely models 1, 2, 3, and 7. The choice of positional uncertainty is not straightforward. It is suggested that, in order to correctly capture lensing systematics (e.g., line of sight structure, complex structure within the cluster), it should be chosen to be large enough that the RMS of each model is comparable to the positional uncertainties assigned to the multiple image systems, and the resulting $\chi^2_\nu$ values are near unity \citep[e.g.,][]{Zitrin2015,Grillo2016}. We stress that this metric is not always a useful guideline (for example, it is not a relevant metric for non-parametric models), and that the choice of positional uncertainty is not agreed upon across lens models in the literature \citep[e.g.,][]{Meneghetti2017,Kelly2023a}. There is also not a consensus whether this guideline is best suited for the case of a single cluster \citep[e.g.,][]{Acebron2022}, whether it should be chosen based off the results of many clusters \citep[e.g.,][]{Zitrin2015}, or drawn from results on simulated clusters \citep[][]{Daloisio2011, Jullo2010}. As seen in Table \ref{tab_lens_info}, all but one lens model (for which chi-squared statistics are applicable) satisfy or exceed these guidelines, implying these models adequately capture or even overestimate their uncertainties. One exception, lens model 3, did show large discrepancies between its RMS and positional uncertainty (e.g., model~3) and by extension a large $\chi^2_\nu$, which warranted further investigation and updating of the choice of positional uncertainty as detailed in Appendix \ref{app:lens_models}. We note that the posteriors of models 2 and 7 are sampled with an effective positional uncertainty of $\sigma_{\rm pos} \sim 1.4\arcsec$, such that their reduced chi-squared's become much lower than one, however this positional uncertainty was found to best account for systematics across a range of clusters \citep{Zitrin2015}.

We emphasize that the choice of positional uncertainty was left to the expertise of each individual lens modeling team. In a landscape where the choice of positional uncertainty is not agreed upon, we favor this approach, and a discussion of appropriate positional uncertainties is outside the scope of this work. However it is interesting to ask how standardizing the models by their positional uncertainty or chi-squared statistics would affect the $H_0$ inference. 
For this exercise, we approximate the posterior of each lens model as a single multivariate Gaussian (this results in only minor information loss since the posteriors are fairly well-behaved), and, assuming a Gaussian likelihood, can approximate the effect of increasing or decreasing the positional uncertainty on the lens model time delay and magnification PDFs as scaling the PDF width linearly with the positional uncertainty (i.e., scaling with $\sqrt{\chi^2_\nu}$).

This was not straightforward for each model; here we detail the modifications made to each model posterior to achieve this standard. For model 1, the posterior width is re-scaled by a factor of $\sqrt{0.44}$ to achieve an effective chi-squared of unity, narrowing the uncertainties by $\sim 33\%$. For lens models 2 and 7, the effective positional uncertainty used to generate the posteriors ($\sigma_{\rm pos}=1.4\arcsec$) was already chosen to have $\chi^2_\nu \sim 1$ across a large sample of lensing clusters, and as such these are left unchanged. We note that the reduced chi-squared is already far below unity assuming $\sigma_{\rm pos}=1.4\arcsec$, and re-scaling the posterior width would result in much smaller uncertainties on the time delays and magnifications. For model 3, which had $\chi^2_\nu=0.30$, the uncertainties are rescaled by a factor $\sqrt{0.30}$. In the case of model 5, positional uncertainties and chi-squared statistics are not applicable; instead an effective positional uncertainty is estimated as $\sigma = 0.18\arcsec$ and the posterior widths are rescaled by $0.1/0.18=0.56$ to match the positional uncertainty to the RMS of $0.1\arcsec$ (this is similar to achieving a reduce chi-squared of 1 when assuming a Gaussian likelihood, as seen in model 1). Models 4 and 6 are left unchanged: model 4 does not make use of positional uncertainty \citep[a term for the mean source scatter is used in uncertainty estimation, however it is not functionally equivalent the positional uncertainty term of typical parametric lens models; see][]{Cha2022, Cha2023}, and model 6's uncertainties are determined via modification of the lensing assumptions, not the positional uncertainty.

Performing this test and recomputing the $H_0$ inference results in only very minor changes: $H_{\rm 0, TD-only}=72.9^{+9.9}_{-8.0}$, $H_{\rm 0, phot-only}=75.3^{+16.6}_{-5.3}$, $H_{\rm 0, phot+spec}=74.9^{+6.5}_{-4.7}$. As is apparent, none of the inferences are affected in a statistically significant fashion, despite most of the lens models having a substantial reduction in their uncertainties. This follows the expectation of Section \ref{sec:disc_err}, where the uncertainty of the individual lens model is subdominant to the uncertainty induced by the spread between lens models.

While the results of this exercise are to be taken with caution, as there is no straightforward way to standardize the diverse set of lens models on the basis of positional uncertainty or chi-squared statistics, it further emphasizes the findings of Section \ref{sec:disc_err}. The precision of the magnification and time delay ratio measurements from photometry and spectroscopy are such that they can find agreement with multiple lens models, however these lens models can exhibit $\gtrsim1\sigma$ tensions between one another. In this regime, the disagreement between lens models is the dominant barrier in reducing uncertainty on $H_0$, highlighting the importance of using a range of models to capture lensing systematics. This in turn implies that the uncertainty estimation for any individual lens model is likely sufficient for this inference, but as time delay and magnification measurements improve (thereby improving the weighting scheme) or as the agreement between lens models improve (e.g., refining multiple image systems, using magnifications or time delay ratios as model constraints), the uncertainty of individual lens models will ultimately have the greatest impact on the $H_0$ uncertainty.

\section{The Plane of Optimization} \label{app:spvsip}
This analysis includes lens models which optimize both in the image plane and in the source plane (see Table \ref{tab_lens_info}). Both approaches have been shown to perform similarly in previous analyses on galaxy cluster lenses \citep[e.g.,][]{Meneghetti2017}, however this has not been thoroughly evaluated for time delays in the literature. More generally, for this precision measurement, it is useful to evaluate the effects of the plane of optimization. Across the seven lens models in the sample, no clear trend emerges upon comparison of the source plane and image plane models, however the approaches of these models are in some cases substantially different. To better quantify this effect, we investigate models 1, 2, and 3, which had the functionality to optimize in the image plane and the source plane (however we note that this functionality is not typically used in the literature for models 1 and 2). Model 4 cannot be optimized in the image plane, however this effect is likely subdominant to the the regularization term used in its minimization \citep[see ][]{Cha2022}.

Each image plane model was run using the same constraints as its source plane counterpart, and the resultant magnifications and time delays were compared. For Models 1 and 2, the source- and image-plane models were found to agree at the $<1\sigma$ level; this corroborates findings from previous testing of these softwares on simulated lenses \citep{Furtak2023MNRAS.523.4568FLensModel, Oguri2010}. By contrast the model 3 source plane model was found to produce larger magnifications by $\sim 10\%$ while inferring similar time delays. In principle, the use of the magnification tensor in the source plane chi-squared should prevent such an effect, preventing models from arbitrarily scaling the magnification to minimize the scatter of images in the source plane \citep[e.g., ][]{Oguri2010}. One caveat is that source plane models may find unphysical solutions, predicting counterimages which are not observed, however upon inspection this does not occur in model 3. We speculate that the software model 3 uses, \texttt{LENSTOOL}, may struggle with posterior estimation with the source-plane $\chi^2$. \cite{Jullo2007} caution this software may be less robust with this approach. Indeed in testing we found the source-plane optimization to behave in unreliable ways (e.g., the choice of positional uncertainty affected the posterior medians) whereas the image-plane optimization did not. Furthermore providing the image-plane model solution to the source-plane $\chi^2$ yielded a better solution than what was found in the source-plane optimization.

Across the three models, this may tentatively suggest that overall source-plane optimization is a sufficient approximation for the current lens model precision. However we caution that for this cannot be taken for granted, and some models show significant deviations in their solutions between the source- and image- plane. As lensing constraints grow tighter (e.g., more spectroscopic redshifts, more image systems), image-plane modeling may be preferred, and further detailed comparison between these approaches is needed in future work.

\section{Probability Distribution Function Parametrization and Likelihood Integration}\label{app:prob}
The probability distribution functions (PDFs) from  the photometric light-curve fitting, spectroscopic fitting, and the lens model fitting are parametrized as Gaussian-mixture models (GMMs). 
Each fit (photometry, spectroscopy, or lens model), $F$, generated a few hundred to thousands of samples from the posterior distribution, $P(\mathcal{O}|F)$, which was then fit by a linear combination of $N$ multivariate Gaussians such that:
\begin{align}
    P(\mathcal{O}| F) \approx \sum^{N}_{i=1} w_i \mathcal{N}_{F}(\mathcal{O} | \mu_i,\mathcal{C}_i)\quad,
\end{align}
where $w_i$ is the weight assigned to each normalized multivariate Gaussian, $\mu_i$ and $\mathcal{C}_i$ are the associated mean vectors and covariance matrices respectively, and $N$ was chosen for each model based on the Bayesian information criterion (BIC) to avoid overfitting to artifacts which may result from undersampling. Thus, the integrand in Eq.~\ref{eq:like_sum} becomes the product of two GMMs or equivalently the sum of products of multivariate Gaussians:
\begin{align}
    \int P(\mathcal{O}|M_l ; H_0) P(\mathcal{O}| {\rm LC})d\mathcal{O}_1 ... d\mathcal{O}_n 
    =  \int \sum^{N}_{i=1}\sum^{N}_{j=1} w_i w_j \mathcal{N}_{M_l}(\mathcal{O} | \mu_i,\mathcal{C}_i) \mathcal{N}_{LC}(\mathcal{O} | \mu_j,\mathcal{C}_j) d\mathcal{O}_1 ... d\mathcal{O}_n\quad,
\end{align}
where $i$ indices refer to the Gaussians which make up the GMM representing the lens model posterior, and the $j$ indices refer to the Gaussians which make up the GMM representing the light-curve fitting posterior. This formalism is  advantageous in that this integral can be solved analytically as the multiplication of two multivariate Gaussians is itself a multivariate Gaussian with a known renormalization.

The combined photometric--spectroscopic likelihood of Eq.~\ref{eq:like_sum_spec}  follows the same formalism, where the posterior from spectroscopic fitting is also parametrized as a GMM, and the integrand is the distributive product of three GMMs. Because the product of any two multivariate Gaussians is itself a Gaussian, the product of 3 multivariate Gaussians is also a Gaussian.

\section{Unblinded Lens Models}\label{app:unblinded}
After the submission of the  blinded model results,
the contributing lens model teams also had the option to construct unblinded lens models, where absolute magnifications measured from light-curve fitting were used as constraints.
Only the teams for models 2 and 4 did so, but these two models represent both parametric and non-parametric types. 
Figure~\ref{fig:lens_corner} illustrates parameter changes and their  covariance for model 2. Reassuringly, the unblinded model 2 posterior follows the expectations from the correlations seen in the blinded model posterior, where the median predicted time delay increases due to the now-lower  magnifications. The non-parametric model 4, however, saw very little change in its time delay predictions. The blinded version of model 4 predicts fairly accurate magnifications with the exception of $\mu_a$, but $\mu_a$ is the only magnification which lacks  significant covariance between the time delays, resulting in only minor changes to the time delays in the unblinded model. 

For both unblinded models, we infer only $H_{0,{\rm TD-only}}$ because magnifications became constraints in the lens models rather than predicted observables. While model 4 gives similar results in both the blinded and unblinded cases, the unblinded model 2 gives a much larger $H_0$ than the blinded model. Interestingly, both unblinded models predict the time delay ratio more accurately than their blinded counterparts, implying they would receive generally higher weighting in the $H_{0,{\rm TD-only}}$ inference. While it is difficult to make strong conclusions based on a sample of only two models, it is interesting to ask whether unblinding the magnifications
may better address systematics between lens models, allowing for more modeling approaches to accurately reproduce the observables and hence contribute to the $H_0$ inference. As the sample of cluster-lensed SNe~Ia grows, functionality for magnification constraints will likely become more ubiquitous in lens modeling techniques, creating an opportunity for a more comprehensive view of lens model systematics. 

\section{Testing Lens Model Systematics} \label{app:systematics}
All seven lens models were fully blinded from the time delays, magnification ratios, and absolute magnifications inferred from the photometric light-curve fitting of \cite{Pierel2023H0pe}. While these measurements provide strong constraining power for weighting model contributions to $H_0$, they also 
introduced an opportunity 
to test lens model systematics using the absolute magnification at the galaxy cluster scale with a multiply-imaged source. 
Note that this is a more stringent test than lower magnification singly-imaged cases such as SN HFF14Tom in the Abell 2744 cluster field \citep{Rodney2016}. Here we broadly follow \cite{Rodney2016} to provide an in-depth discussion of the performance of the contributing models, which in turn determines the weighted contributions to the $H_0$ measurement. 

\textit{Parametric versus Non-Parametric}:
While the contributing models are broadly divided into two categories, parametric and non-parametric, each model is distinct from one another.
Here we compare the most free-form models, models 4 and 6, to the three classically parametric models, models 1, 2, and 3.

We find the non-parametric models have longer relative time delays 
(assuming the fiducial $H_0$ value),
and hence higher $H_0$ inferences than the parametric models. The inferred $H_0$'s also have higher uncertainties.
This may be owing to the greater flexibility of non-parametric models, which can explore solutions for the lens potential inaccessible to rigid fully-parametric profiles. Owing to the lack of sufficient lensing constraints about the northeastern core of the cluster, parametric methods do effectively reduce the parameter space by assuming the analytic profiles are centered at the locations of the bright cluster galaxies. 
As shown in Table~\ref{tab_h0}, both parametric and non-parametric models receive significant weighting under all three inferences, where the weighting reflects the agreement between model predictions of observables and their corresponding measurements from photometry and spectroscopy.
As the non-parametric models tended to predict longer time delays, this emphasizes the need for a wide variety of lens model approaches to fully capture the range of possible solutions.

\textit{Strong versus Strong+Weak}:
Model 4 makes use of both weak lensing and strong lensing constraints. 
While only this model was used for the $H_0$ inference, 
an additional version of the model using only strong lensing was also constructed for testing purposes only. This comparison offers a rare opportunity to evaluate the benefits weak lensing. 
The addition of weak lensing constraints are found to generally increase the time delay uncertainties without greatly impacting their median, while decreasing uncertainties on the magnifications and generally shifting their median lower. This significantly improves the agreement of both the lens model-predicted time delay ratios and the predicted magnifications with those measured by photometry and spectroscopy. As a result, the model receives a much greater weighting in the $H_{0,{\rm phot+spec}}$ inference, increasing the weighting to 14\% for strong+weak constraints rather than $<1\%$ for strong-lensing alone. This suggests that weak lensing constraints are crucial for non-parametric modeling methods, especially when the available number of strong-lensing constraints is low.

\begin{deluxetable*}{cccccccc}
\tabletypesize{\footnotesize}
\tablecaption{Lens model Information}
\tablecolumns{8}
\tablehead{
\colhead{\bf\#} &   \colhead{{\bf Model and Reference}} & \colhead{\bf RMS~($^{\prime \prime}$)} & \colhead{\bf $\sigma_{\rm pos, SN}$($^{\prime \prime}$)} &  \colhead{\bf $\sigma_{\rm pos, global}$($^{\prime \prime}$)} & \colhead{\bf DoF} & \colhead{\bf $\chi_\nu$} & \colhead{\bf Optimization}} 
\startdata
1 & \texttt{GLAFIC} \citep{Oguri2010,Oguri2021} & $0.07$ & $0.1$ & $0.1$ & $119$ & $0.44$ & Source \\
2 & \texttt{Zitrin-analytic} \citep{Zitrin2015} & $0.21$ & $0.1$ & $0.3^{a}$ & $86$ & $0.61$ & Source \\
3$^b$ & \texttt{LENSTOOL} \citep{Kneib2011b} & $0.16$ & $0.3$ & $0.3$ & 109 & 0.3 & Image \\
4 & \texttt{MARS} \citep{Cha2022} & $0.07$ & N/A & N/A & N/A & N/A & Source \\
5 & \cite{Chen2020} & $0.1$ & $0.18{^c}$ & $0.18{^c}$ & N/A & N/A & Image \\
6 & \texttt{WSLAP+} \citep{Diego2005} & $1.1$ & $0.03$ & $0.03$ & N/A & N/A & Image \\
7 & \texttt{Zitrin-LTM} \citep{Zitrin2009} & $0.7$ & $0.1$ & $0.5^{a}$ & $95$ & $1.23$ & Image \\ 
\enddata
\tablecomments{Columns are the RMS separation between model and observed positions lensed images (RMS), the positional uncertainty for multiple image systems used for the $\chi^2$ minimization ($\sigma_{\rm pos, global}$), the positional uncertainty used for the SN images ($\sigma_{\rm pos, SN}$), the number of degrees of freedom (DoF) in the model, the $\chi^2$ per DoF ($\chi^2_\nu$), and which plane the minimization was performed in (Optimization). All quantities are given for the best-fit model.  RMS and $\sigma_{\rm pos}$ refer to the image plane unless otherwise specified. `N/A' indicates quantities that are not applicable to non-parametric models, for which   chi-squared statistics are not relevant for parameter estimation.}
\tablenotetext{a}{Uncertainties are extracted using an effective positional uncertainty $\sigma_{\rm pos,global}\sim 1.4^{\prime \prime}$ following the prescription of \cite{Zitrin2015}.}
\tablenotetext{b}{This model was updated following unblinding, the blinded version of the model was run with source plane optimization using a positional uncertainty $0.03\arcsec$, resulting in an RMS $0.26\arcsec$ and $\chi^2_\nu = 72$. This blinded model is detailed in \ref{app:lens_models}.}
\tablenotetext{c}{Effective positional uncertainty approximated for reference; model is not optimized using chi-squared statistics.}
\label{tab_lens_info}
\end{deluxetable*}
\end{appendix}
\end{document}